\documentclass[journal]{IEEEtran}
\IEEEoverridecommandlockouts
\usepackage{cite}
\usepackage{amsmath,amssymb,amsfonts}
\usepackage{algorithmic}
\usepackage{subcaption}
\usepackage{graphicx}
\usepackage{textcomp}
\usepackage{booktabs}
\usepackage{makecell}
\usepackage{xcolor}
\usepackage{url}
\usepackage{lipsum}
\newcounter{remarkcount}
\newcommand{\remarknew}[1]{%
    \stepcounter{remarkcount}
    \textbf{Remark \arabic{remarkcount}:} #1
}
\newcommand{\revise}{\color{black}}

\usepackage{hyperref}

\hypersetup{hidelinks}

\def\BibTeX{{\rm B\kern-.05em{\sc i\kern-.025em b}\kern-.08em
    T\kern-.1667em\lower.7ex\hbox{E}\kern-.125emX}}
\begin{document}

\title{\LARGE\bf Model Predictive Control for Unlocking Energy Flexibility of Heat Pump and Thermal Energy Storage Systems: Experimental Results
}

\author{Weihong Tang$^{a,1}$, Yun Li$^{a,1}$, Shalika Walker$^b$, Tamas Keviczky$^a$
\thanks{$^a$W. Tang, Y. Li and T. Keviczky are with the Delft Center for Systems and Control, 
        Delft University of Technology,  Delft, the Netherlands.
        {\tt\small ttonytany@gmail.com; y.li-39@tudelft.nl; T.Keviczky@tudelft.nl}}
\thanks{$^b$S. Walker is with the Kropman BV, the Netherlands.
{\tt\small shalika.walker@kropman.nl}}
\thanks{$^1$The first two authors contribute equally to this work.}
\thanks{Corresponding author: Yun Li}
\thanks{This work was supported by the Brains4Buildings project under the Dutch grant 
programme for Mission-Driven Research, Development and Innovation (MOOI).}
}

\maketitle

\begin{abstract}
Increasing penetration of renewable energy sources (RES) and electrification of energy systems necessitates the engagement of demand-side management (DSM) to help alleviate congestion in electricity grid. Heat pump and thermal energy storage (HPTES) systems, being energy efficient solutions, are becoming popular in modern buildings and are promising to contribute to demand-side management (DSM) due to their significant share in household electricity consumption. For typical HPTES systems, this paper presents a systematic design framework covering a control-oriented modeling process and energy-flexible model predictive control (MPC) design. The proposed MPC-based DSM strategy offers an innovative solution for efficient DSM by following a two-step DSM framework. In the first step, \textit{flexibility assessment} is performed to quantitatively evaluate the flexibility potential of the HPTES system by solving a mixed-integer economic MPC problem. In the second step, \textit{flexibility exploitation} is achieved through reacting to feasible demand response (DR) requests while respecting system constraints. Both numerical simulations and real-world experiments are performed based on a real HPTES installation to showcase the viability and effectiveness of the proposed design.
\end{abstract}

\section{Introduction}
In pursuit of carbon neutrality and energy sustainability, all economies have set to improve the utilization of renewable energy sources (RES), such as solar energy, wind energy, etc. For example, the European Union (EU) has set an ambitious target to increase the share of RES to at least 27\% by 2030 \cite{dutchgov2022}. It should be highlighted that the intermittency and volatility of RES pose significant challenges for maintaining a real-time balance between electricity supplement and consumption in power grids. With the aim of reducing congestion, the concept of demand-side management (DSM) has been proposed. Unlike conventional supply-side management, DSM involves strategic adjustments of the energy consumption of end users in response to the need of power grids \cite{bunning2022robust, Abl:45}. 

Buildings, as major energy consumers, contribute to about 40\% of total energy consumption. The heating, ventilation, and air conditioning (HVAC) devices consume more than half of the total household energy consumption. Due to the significant energy consumption of HVAC systems in buildings, there has been growing interest in leveraging their energy flexibility to deliver DSM services. As energy-efficient heating/cooling devices, heat pumps (HPs) have gained increasing popularity. The EU's directive on renewables identified heat pumps as a vital technology for exploiting renewable energy \cite{EU2009}. In recent years, HPs are widely adopted across Europe to provide thermal energy for various building applications, such as domestic hot water usage and floor heating. Currently, 60 million heat pumps are installed in Europe, and the number of HP installations is experiencing a sharp rise. In 2022 alone, there were about 3 million HPs installed, and it is expected that at least 10 million additional heat pumps will be installed by 2027 \cite{hp2022}. A typical application of HP is its integration with thermal energy storage (TES) tanks, referred to as heat pump and thermal energy storage (HPTES) systems in this paper. By buffering hot water in water tanks, the integration of TES allows HP to operate flexibly for improving energy efficiency and reducing energy costs without sacrificing the required hot water supply \cite{ermel2022thermal}. Given the high power demands created by HPs, it is promising to explore the possibility of exploiting the energy flexibility emanating from HPTES systems to achieve DSM.

In practice, the majority of the HPTES systems operate using rule-based control strategies: HP is turned on when the TES temperature drops below a specified threshold and turned off when it exceeds that threshold.
While rule-based approaches are easy to implement, they fail to exploit the energy flexibility of the system. It is commonly accepted that advanced control strategies are necessary to harness the energy flexibility of HPTES systems. As an advanced control technique, model predictive control (MPC) has become a promising approach in operating HPs and TES systems in building sectors due to its versatility in dealing with system constraints and in incorporating economic factors and predicted system behaviors, see \cite{kuboth2019experimental,d2019model,kajgaard2013model,mugnini2024model,bunning2022robust, touretzky2014integrating, tarragona2022analysis} and references therein.

{\revise For implementing MPC schemes, prediction models of HP and TES systems are necessary. While there are many existing works developing white-box models of HP and TES systems to give high approximation accuracy, see \cite{kinab2010reversible,shin2004numerical,BAETEN2016217} and the references therein, these models are too complex to be considered in MPC design. To balance the approximation accuracy and computational cost, we develop simplified models for MPC design. In general, HP models can be simplified as COP functions to bridge the relationship between the HP's thermal generation and electricity consumption \cite{tarragona2022analysis,vrettos2016robust,golmohamadi2021optimization,de2016quantification}. As for TES systems, they are typically modeled as either a single thermal node, assuming the water within the tank is properly mixed, or as multiple thermal nodes, neglecting the temperature gradient along radial directions \cite{touretzky2014integrating,d2019mapping}.   
}

{\revise  
Although the existing literature extensively investigated the viability and effectiveness of MPC in controlling HVAC devices in building sectors, including HP and TES systems, there is still room for improvement, particularly in the context of DSM. For the MPC-based DSM design within the context of TES systems, most of the existing approaches adopt either the price-based or incentive-based programs, in which HVAC devices are incentivized by some price signals to voluntarily adjust their operation patterns in line with power grid requirements \cite{farrokhifar2021model,golmohamadi2022integration,d2019mapping}, relying only on unidirectional communication from the grid operator to the resource manager. However, as noted in \cite{li2023unlocking,S22023,konsman2020unlocking}, these approaches, despite being relatively straightforward and easy to implement, often fail to fully exploit the energy flexibility of the system and may not meet the grid's energy reduction goals. This is largely due to insufficient information exchange regarding the system's flexibility potential and expected energy consumption patterns between BMS and grid operators.

The increased penetration of distributed energy sources as well as the electrification of smart building systems enable more advanced DSM strategies to better harness the energy flexibility of energy consumers by establishing privacy-secured information sharing among energy consumers and producers. In order to cater to diverse household devices and different markets and contractual constraints, the European Commission's M/490 mandate has introduced a future-proof energy flexibility interface (EFI) -- the S2 standard -- for leveraging energy flexibility in the built environment \cite{S22023,konsman2020unlocking}. Within the S2 standard, a bi-directional communication flow between the customer energy manager and the resource manager is developed to facilitate the exchange of information on flexibility capabilities and deployment, which paves the way to a versatile and privacy-secured energy flexibility exploitation architecture.

While there are some recent approaches compatible with the S2 protocol for better exploiting the energy flexibility of HVAC in buildings (e.g., \cite{li2023unlocking,li2023robust,bunning2022robust}), these results can hardly be generalized to ON-OFF type HPs with nonlinear system dynamics, which is the case investigated in this paper. Another limitation hindering the practical application of MPC in HPTES systems is the lack of general modeling approaches to develop control-oriented models for typical HPTES systems with balanced model complexity and approximation accuracy. 
Only specific HPTES configurations with tailored control-oriented models are usually considered in most of the existing literature.}

{\revise Motivated by the above discussion, by extending our preliminary work in \cite{10666588}, this paper presents a comprehensive approach to MPC design for HPTES systems, including modeling, control algorithm design, numerical simulation and real-world implementation, with the aim to harness the energy flexibility of HPTES systems to mitigate grid congestion issues without degrading system performance. Our key contributions are as follows.
\begin{itemize}
    \item General control-oriented models for the main components of HPTES systems -- heat pump, thermal tanks and hot water consumption -- are introduced that strike a balance between model complexity and prediction accuracy, and are adaptable to various alternative HPTES configurations. Possible extensions of the introduced models are also discussed.
    \item An energy-flexible MPC design is proposed for the HPTES system. Notably, based on classical economic MPC formulations, a two-step DSM scheme is developed to better harness the energy flexibility of the system. By imposing a set of mixed-integer linear constraints, our proposed approach can quantitatively evaluate the energy flexibility of the HPTES system regardless of the explicit system model. The energy flexibility can also be safely exploited via DR request mechanisms.    
    \item Both numerical simulation and real experiments are conducted based on a real-world HPTES installation to validate the viability and effectiveness of the proposed approach. Practical solutions for improving the computational efficiency of MPC are discussed. The results show that adopting the proposed MPC schemes not only leads to reduced energy cost, but also enables a reliable assessment and exploitation of the energy flexibility of the HPTES system while respecting system constraints.
\end{itemize}
}

The remaining parts of this paper are organized as follows. Section II presents a control-oriented model for the HPTES system. Section III introduces the MPC design framework for both the economic operation and DSM of the HPTES system. Numerical and experimental results are detailed in Section IV. Finally, Section V concludes the paper.

\section{Control-Oriented Modeling for HPTES System and Problem Formulation}\label{sec:models}

As stated in the introduction, an MPC-based DSM scheme will be designed for the HPTES system, where a prediction model for the HPTES system is used. In this section, a general control-oriented model of HPTES systems will be introduced, accompanied with the system constraints, as well as a prediction model for the hot water consumption.

\subsection{Modeling for HPTES System}
\begin{figure}
\centering
    \includegraphics[width = \linewidth]{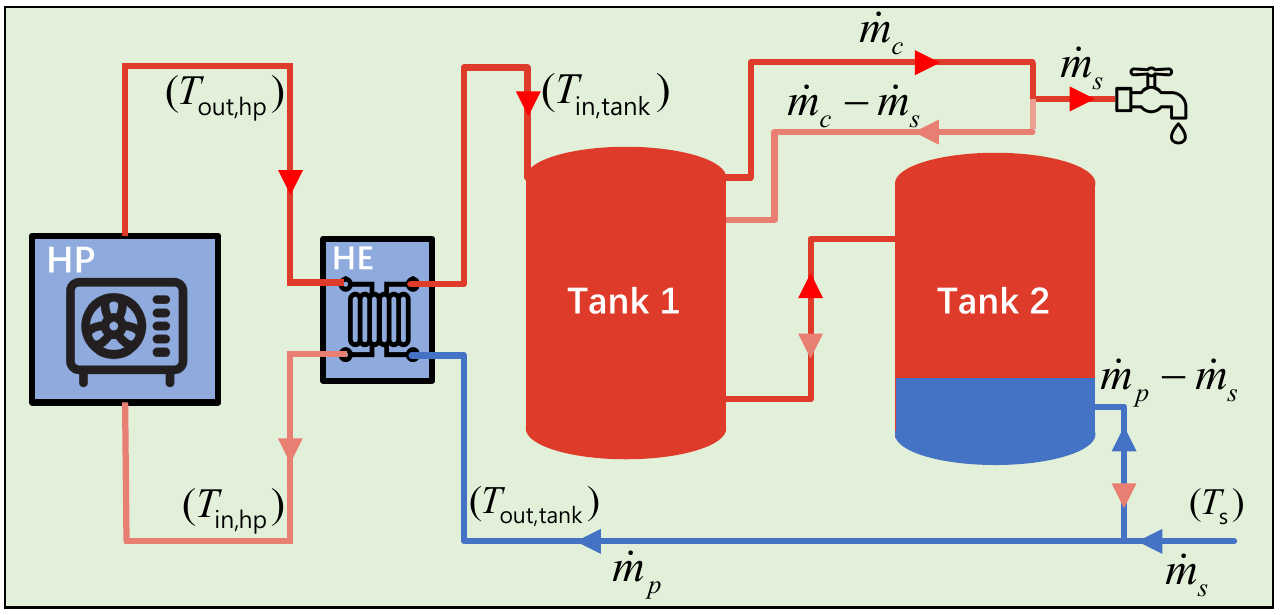}
    \caption{Diagram of the considered general HPTES system \cite{10666588}.}
    \label{fig:HPTES}
\end{figure}

 Based on a real HPTES installation in an office building in the Netherlands, Fig. \ref{fig:HPTES} depicts a general configuration of an HPTES system, where red lines denote hot water circulation, and blue lines denote cold water circulation with arrows indicating the direction of water flows. 
 In this system, hot water from the top of Tank 1 is circulated at a mass flow rate $\dot{m}_c$ through connected taps to meet hot water usage demands. The real-time hot water consumption, denoted by $\dot{m}_s$, is assumed to be predictable with sufficient accuracy based on historical data (our proposed method for this is discussed in Section \ref{ch_pred_water}). Whenever hot water is consumed, an equal amount of cold water $\dot{m}_s$ will be simultaneously supplied to the bottom of Tank 2, ensuring that the total water volume within the HPTES system remains constant. If the HP is off, the cold water is directly injected into the bottom of Tank 2. Conversely, when the HP is on, water from the bottom of Tank 2 with a mass flow rate $\dot{m}_p$ will be transmitted to the top of Tank 1 after being heated through the heat exchanger.

Such an HPTES system has two major components: heat pump and thermal energy tanks, which are important in designing an MPC-based DSM scheme. In the following subsections, we will provide general control-oriented models for the HP and the thermal tank, respectively, which will be utilized for our MPC design in Section \ref{sec:mpc}.

\subsubsection{Modeling for Heat Pump}
 HPs are responsible for providing thermal energy to the HPTES system by consuming electricity. Accurately modeling the relationship between electricity consumption and thermal energy generation is crucial for designing control strategies to improve the system's energy efficiency and safety. However, a detailed and precise white-box HP model can be complicated and also requires an extensive understanding of the interval working principles of each component of HPs, which demands expert knowledge and will also make the control problem computationally intractable \cite{correa2018air}.  
 To balance the system complexity and computational burden, we adopt a grey-box model -- coefficient of performance -- to build the relationship between the thermal energy output and the electricity consumption of HPs.
 
{\revise The coefficient of performance (COP) is defined as the ratio of heat output to work input \cite{wang1983heat}. Mathematically, it is expressed as:
\begin{equation}\label{eq:cop}
    \text{COP} = \frac{Q_{\text{hp}}}{P_{\text{hp}}},
\end{equation}
where $Q_{\text{hp}}$ denotes the heat output of the HP and $P_{\text{hp}}$ its power input. Since the COP generally varies with different operation conditions in a nonlinear relationship, it is approximated with a function $f_{\text{COP}}(\cdot)$ in MPC design. Besides, the HP in our HPTES system is assumed to operate in two modes: on and off, which is common in many practical installations, so that the HP power input is $P_{\text{hp}} = P_ru$ with $P_r$ as the rated HP power and $u\in\{0,1\}$ the on/off HP control signal. Consequently, the thermal energy generated by the HP can be represented as:
\begin{equation}\label{eq:hp}
    Q_{\text{hp}} = f_{\text{COP}}(\cdot) P_{r} u,
\end{equation}
where $f_{\text{COP}}(\cdot)$ is an function approximating the COP values under different operation conditions.

The COP is influenced by HP's working conditions, and two main factors are the supply water temperature (condenser outlet) and the ambient heat source temperature \cite{tarragona2022analysis}. To balance the trade-off between approximation accuracy of the model and the computational complexity of the resulting MPC problem, the following bilinear COP model is utilized
\begin{equation}\label{eq:cop2}
f_{\text{COP}}(T_{\text{in,hp}},T_{\text{amb}}):=a_1+a_2 T_{\text{in,hp}}+a_3  T_{\text{amb}}+a_4  T_{\text{in,hp}}  T_{\text{amb}},
\end{equation}
where $T_{\text{in,hp}}$ is the inlet water temperature of HP, $T_{\text{amb}}$ is the ambient air temperature of HP, and $(a_1,a_2,a_3,a_4)$ are parameters to be identified. Similar COP approximation can also be found in \cite{felten2018value,tarragona2022analysis}.
}

Compared with the constant COP model, which is widely utilized in existing literature \cite{luickx2008influence}, the bilinear model in \eqref{eq:cop2} captures the major factors of COP -- the heat source (ambient air) and the heat consumer (inlet water to be heated) -- that influence the HP performance to provide a more practical and accurate model without demanding high computational burden. 

\subsubsection{Modeling for Thermal Energy Storage}
For the HPTES system, thermal tanks can improve the system's energy efficiency and flexibility by storing hot water. By buffering hot water, the HP can operate more flexibly regardless of the real-time hot water consumption. There are various approaches for modeling the dynamics of thermal storage tanks, with different emphases on model accuracy and simplicity. A three-dimensional model, which is generally based on computational fluid dynamics to capture the detailed internal thermal behaviors, can achieve high approximation accuracy but is computationally expensive \cite{BAETEN2016217}. By using a fully mixed tank model or lumped parameter model, the entire water tank is simplified as a single thermal node without considering any internal dynamics \cite{bunning2022robust}. While this mixed tank model is computationally simple, the prediction accuracy is unsatisfactory, especially for large water tanks where internal dynamics cannot be neglected. To balance model fidelity with computational efficiency, a one-dimensional model only considers the vertical internal thermal interactions within the water tank \cite{DELACRUZLOREDO2023120556,rastegarpour2018predictive}. 
\begin{figure}[h!]
            \centering
    \includegraphics[width=0.95\linewidth]{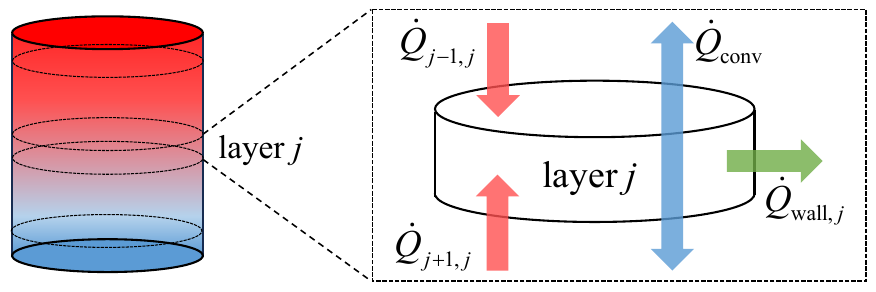}
            \caption{Heat flow diagram of a single layer of TES.}
            \label{fig:node}
\end{figure}
In our work, a one-dimensional stratified water tank model is adopted to approximate the thermal dynamics of water tanks. We divide each water tank into several stratified layers with each layer being modeled as a single thermal node with a uniform temperature. For adjacent water layers, the thermal interactions are considered. The general thermal behavior of each layer is illustrated in Fig. \ref{fig:node}. For each thermal layer $j$, its thermal dynamics are driven by three components:
\begin{itemize}
    \item $\dot{Q}_{j+1,j}$ and $\dot{Q}_{j-1,j}$: thermal conduction between layer $j$ and its adjacent layers $j+1$ and $j-1$.    
    \item $\dot{Q}_{\text{conv}}$: thermal convection due to water flows $\dot{m}_t$ in/out of $j$-th layer from/to adjacent layers.
    \item $\dot{Q}_{\text{wall},j}$: thermal loss through the tank wall in layer $j$.
\end{itemize}

The thermal dynamics of each thermal node can be expressed by the following equations:
\begin{subequations}\label{eq:tank_model}
\begin{align}
& m_{j} c_p \frac{dT_j}{dt}=-\dot{Q}_{\text {wall}, j}+\dot{Q}_{j+1,j}+\dot{Q}_{j-1,j}+\dot{Q}_{\text {conv}}, \\
& \dot{Q}_{\text {conv}}\!=\!
\begin{cases}\dot{m}_t c_{p}(T_{j-1}-T_{j}),\ \dot{m}_t\text{ flows from $j\!-\!1$ to $j$ } \\
\dot{m}_tc_p(T_{j+1} -T_j),\ \dot{m}_t\text{ flows from $j\!+\!1$ to $j$}
\end{cases}\\
& \dot{Q}_{j+1,j}=R_{j,j+1}\left(T_{j+1}-T_{j}\right), \\
& \dot{Q}_{j-1,j}=R_{j-1,j}\left(T_{j-1}-T_{j}\right) , \\
& \dot{Q}_{\text {wall},j}=R_{w,j}\left(T_{j}-T_{\text{amb}}\right),
&
\end{align}
\end{subequations}
where $m_j$ is the mass of the $j$-th node, $c_p$ is the specific heat capacity of water, $\dot{m}_t$ is the water mass flow rate between adjacent tank layers, $R_{j,j+1}$ and $R_{w,j}$ are thermal resistances between the $j$-th layer and the $(j+1)$-th layer as well as the tank wall, respectively.

{\revise Equation \eqref{eq:tank_model} describes the general thermal behavior of a water layer. Special consideration is needed for layers with hot water extraction, i.e., the top layer of Tank 1, and cold water supply, i.e., the bottom layer of Tank 2. For the top layer of Tank 1, which does not have an upper layer, the thermal conduction term $\dot{Q}_{j-1,j}$ is zero, and the thermal convection term $\dot{Q}_{\text{conv}} = \dot{m}_t c_p (T_{j-1} - T_j)$ is computed with $T_{j-1}$ as the temperature of the inlet hot water entering Tank 1 from the heat exchanger, and the water mass flow rate $\dot{m}_t$ is set as the water circulation mass flow rate $\dot{m}_p$. A similar analysis applies to the bottom layer of Tank 2, where no lower layer exists but with a cold water supplement. 
In addition, for our HPTES system, since the direction of water circulation changes with HP operation status (indicated by the control signal $u$), the internal mass flow rate of water convection $\dot{m}_t$ is a function of the HP control signal $u$ as well as the mass flow rate of hot water consumption $\dot{m}_s$.

Furthermore, the water pumps driving the water flow of cold water supply $\dot{m}_s$ and water circulation within tanks $\dot{m}_p$ in \eqref{eq:tank_model} are assumed uncontrollable in our design, which is the case for many practical HPTES installations. It is worth pointing out that the control-oriented model \eqref{eq:tank_model} can be extended to consider these variables as controllable terms, which can then be optimized in the MPC design of Section \ref{sec:mpc}, without changing the model structures. This can be expected to improve system efficiency, but would also increase the computational cost for solving the corresponding MPC problem by inducing extra nonconvex terms. Consequently, this choice should be made based on the hardware availability and the trade-off between expected control performance and computational cost.
}

\remarknew{The HPTES configuration along with the control-oriented model introduced in this section is sufficiently general to cover a wide range of HPTES systems and is flexible enough to be adjusted to cater to other alternative configurations with only minor modifications. For instance, the positions of the hot water extraction and the cold water supplement, as well as the internal water circulation directions can be flexibly adjusted without breaking the structure of the control-oriented model. Furthermore, this model is computationally tractable to be incorporated into MPC design as a prediction model.}

\subsection{System Constraints}
To ensure operational efficiency and safety, the HPTES system has to respect several physical constraints. In this subsection, several typical constraints will be introduced for the HPTES system.

Firstly, the water temperature in the water tanks should be maintained within an admissible range to ensure safety and to provide qualified hot water. The water temperature constraint can be reformulated as the following linear constraint
\begin{equation}\label{eq:temp_cons}
    x^l \leq x \leq x^u,
\end{equation}
where $x$ denotes the temperature vector of different water layers and pipelines, $x^l$ and $x^u$ are the corresponding temperature lower bound vector and upper bound vector, respectively. It should be highlighted that, among all water layers within the water tanks, the water temperature at the top layer of Tank 1 is of more importance since it is more related to the water to be supplied to the buildings.

Furthermore, frequent on-off switching of the HP should be prevented as it can lead to excessive wear and tear, reducing the HP's operational lifespan. Consequently, the number of on-off switches within a time period should not exceed a threshold. The constraints for the number of switches can be formulated as the following quadratic constraint
\begin{equation}\label{eq:switch}
\sum_{k=1}^{m}(u_{t-k+1}-u_{t-k})^2 \leq N_{\text {ctrl}},
\end{equation}
where $N_{\text{ctrl}}$ is the maximal number of switches that is allowed during $m$ time steps, and $u_{t-k}$ denotes the control input at the time instant $t-k$ (i.e., $k$ time steps prior to the current time instant $t$). Since $u_t\in\mathbb{B}$, the left-hand side of \eqref{eq:switch} computes the number of switches that occur during the most recent $m$ time steps. For instance, setting $m=8$ and $N_{\text {ctrl}}=1$ implies that, only one switch is allowed within an 8-step time window.

\subsection{Prediction of Water Consumption}\label{ch_pred_water}

Accurate prediction of the water consumption is important for improving the performance of MPC-based control strategies since the amount of consumed hot water will directly influence the thermal dynamics of the thermal tanks, and consequently the operation pattern of the HP. This subsection introduces a prediction model of hot water consumption via the seasonal autoregressive integrated moving average (SARIMA) model \cite{dubey2021study,box2015time} to effectively capture the inherent water consumption patterns. 

The ability of SARIMA model to incorporate both non-seasonal and seasonal elements makes it a simple and robust tool for forecasting in contexts with repetitive patterns, which is often the case for hot water consumption in HPTES systems. While there are other state-of-the-art prediction schemes available with extraordinary performance, such as artificial neural networks, the limited historical data in our specific real life system limits the applicability of these prediction schemes. Besides, developing high-performance hot water prediction models is not our focus in this paper. Thus, We adopt this SARIMA model as an example to provide hot water predictions. 
 
\begin{figure}
            \centering
            \includegraphics[width = \linewidth]{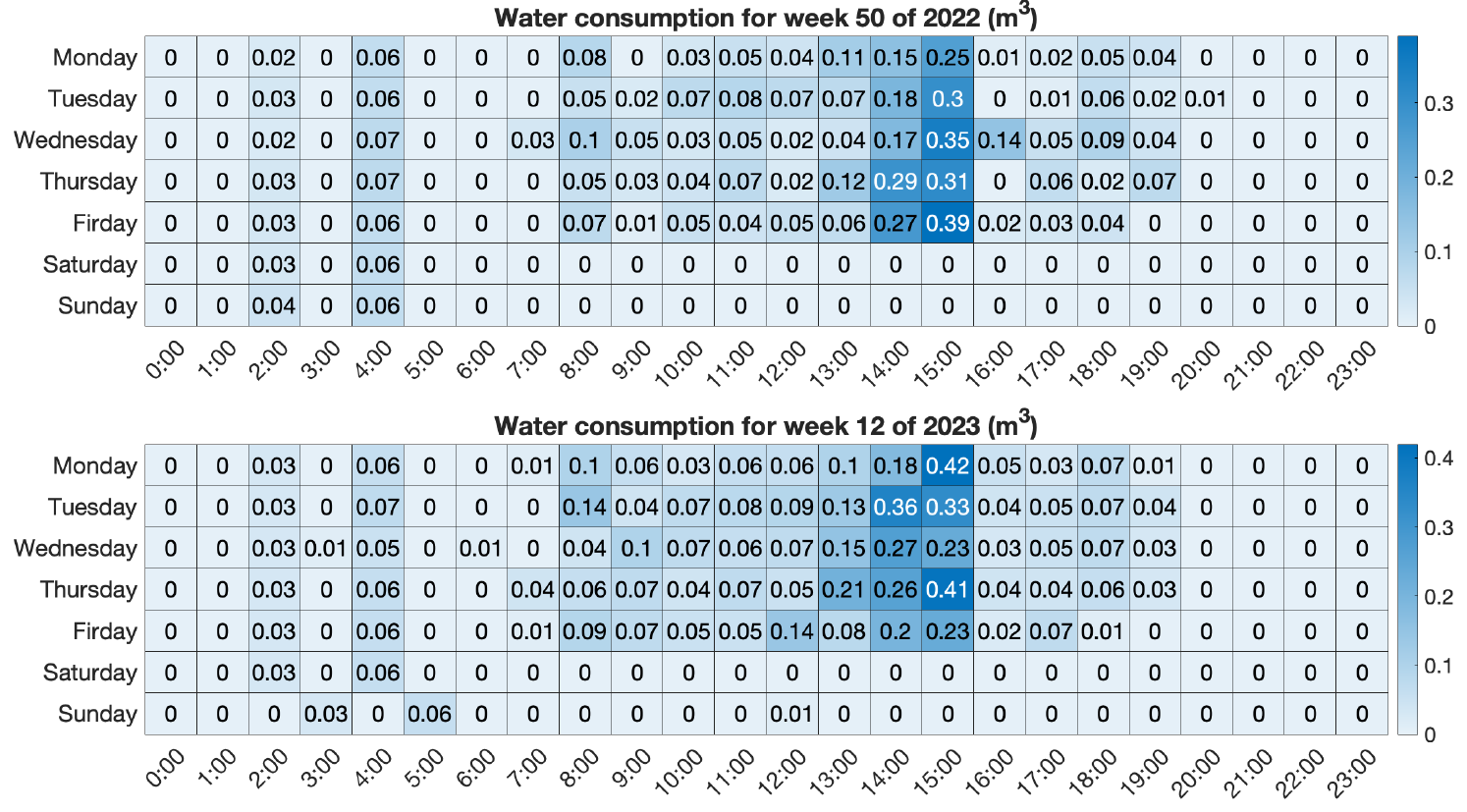}
            \caption{Heat map for water consumption}
            \label{fig:heatmap_water}
\end{figure}

As shown in Fig. \ref{fig:heatmap_water}, the water consumption patterns for our real HPTES installation are similar for different days, where peak consumption happens around 15:00 each workday. To exploit this daily pattern, the SARIMA model is adopted for hot water consumption forecasting in our work. 

A typical seasonal autoregressive integrated moving average (SARIMA) model, denoted as $\text{SARIMA}_{(p, d, q)\times(P, D, Q,S)}$, can be expressed as
\begin{equation}
\begin{aligned}
&(1-\sum_{i=1}^p\phi_i L^i)(1-L)^d(1-\sum_{i=1}^P\Phi_i L^{iS})(1-L^S)^D{Y}_t  \\
&=(1+\sum_{i=1}^q\theta_i L^i)(1+\sum_{i=1}^Q\Theta_i L^{iS})\varepsilon_t
\end{aligned}
\end{equation}
where:
\begin{itemize}
    \item $L$ is the lag operator, i.e., $L^ky_t=y_{t-k}$.
    \item ${Y}_t$ is the hot water consumption at time instant $t$.
    \item $\varepsilon_{t}$ is the white noise error representing the forecast error at time instant $t$.
    \item $S$ is the seasonality period representing the number of time steps in one seasonal cycle.
    \item $(\phi_i,\Phi_i,\theta_i,\Theta_i)$ are coefficients to be identified.
    \item $p$ and $P$ represent the order of the past autoregressive non-seasonal terms and seasonal terms, respectively.
    \item $d$ and $D$ denote the order of non-seasonal and seasonal differencing terms, which are used to make the time series stationary by removing trends or seasonality.
    \item $q$ and $Q$ denote the order of the non-seasonal and seasonal moving average (MA) terms, respectively, which represent how many lagged forecast errors are used to predict the current value.
\end{itemize}
For our case study, for example, the seasonality parameter $S$ is selected to achieve a 24-hour seasonality to exploit the daily pattern of the water consumption.

It should be pointed out that the MPC design proposed in this work does not require a specific prediction model of hot water consumption. Any models that can provide water consumption prediction can fit in our proposed control strategy.

\section{MPC Design for Economic Operation and Demand-Side Management}\label{sec:mpc}
{\revise In this section, based on the two-step DSM protocol, which is introduced in \cite{li2023robust,li2023unlocking} aligning with the S2 standard introduced by the European Commission for exploiting 
energy flexibility in the built environment \cite{S22023}, an energy-flexible MPC design framework is proposed for the HPTES system to achieve economical operation and DSM. 

Fig. \ref{fig:simp_imple} provides a comprehensive workflow for our proposed MPC framework to achieve economic operation and DSM with HPTES systems, which consists of two main strategies: 1) economic MPC, and 2) DSM-based MPC. When no DR requests are activated, the HPTES system is operated with the economic MPC strategy for reducing the energy cost. Once the DR requests are detected, the DSM-based MPC strategy will be deployed to provide energy flexibility services. Once the DR request is finished and no further DR requests are in the queue, the control scheme will be switched back to the economic MPC strategy.
\begin{figure}[h]
    \centering
   \includegraphics[width=\linewidth]{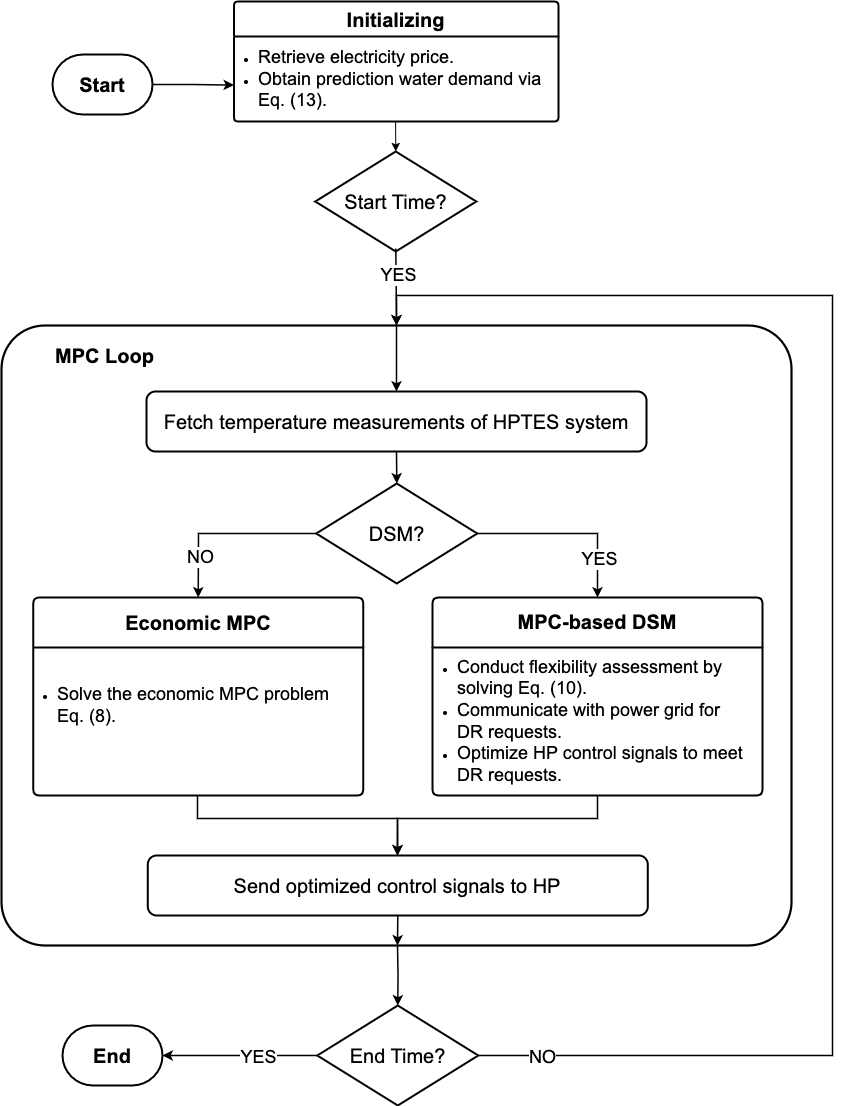}
    \caption{Workflow for implementing the proposed MPC schemes.}
    \label{fig:simp_imple}
\end{figure}}

For the HPTES system, the economic MPC formulation aiming at reducing the energy cost of the HPTES system is expressed as the following optimization problem
\begin{subequations}\label{eq:economic_mpc}
    \begin{align}
        \min_{u_t}\ & J_o \\
        \text{s.t. } & x_{t+1} = f(x_t,u_t,d_t) \\
        & (x_t,u_t) \text{ satisfy } \eqref{eq:temp_cons}\ \&\ \eqref{eq:switch},\\
        & \text{for } t = 0,1,\cdots,N
    \end{align}
\end{subequations}
where $J_o := \sum_{t=0}^Nl(x_t,u_t)$ is the total operational cost function with $l(x_t,u_t)$ as the stage cost and $N$ the length of the prediction horizon, $x_t$ denotes the system states composed of the temperature vector of different water layers and pipelines, $u_t\in\mathbb{B}$ is the HP binary control signal with $u_t = 1$ indicating on status, $f(x_t,u_t,d_t)$ is the control-oriented model introduced in \eqref{eq:hp} and \eqref{eq:tank_model}. In our design settings, the operational cost for the HPTES system mainly comes from electricity consumption. Consequently, the operational cost function is defined as $J_o:= \sum_{t=0}^N e_t\cdot P_r \cdot u_t$, where $e_t$ is the electricity price signal and $P_r$ is the rated power of HP.

\begin{figure}
    \centering
    \includegraphics[width = 0.95\linewidth]{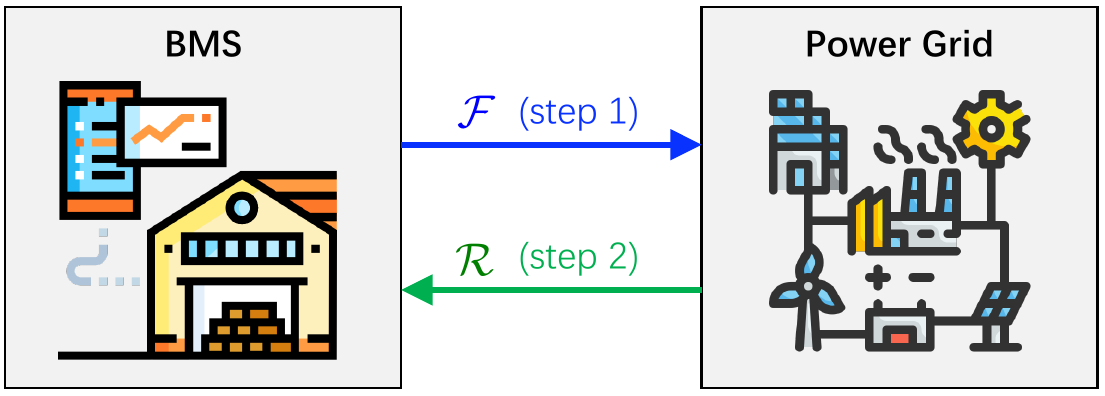}
    \caption{Diagram of the two-step DSM scheme \cite{10666588}.}
    \label{fig:dsm}
\end{figure}

Based on the above economic MPC formulation, in the following, we will show how to further enable energy use flexibility in a DSM scheme. Current mainstream DSM design follows the so-called price-based or incentive-based programs, which are one-step programs and can be inefficient in exploiting the energy flexibility of the system. Instead of following the conventional design framework, our work proposes a novel DSM framework based on a bidirectional communication protocol between the building management system (BMS) and the grid operator to achieve better utilization of energy flexibility. Fig. \ref{fig:dsm} shows the diagram of the DSM framework, which consists of two steps.
At \textit{Step 1}, called \textit{Flexibility Assessment}, the energy flexibility potential of the HPTES system is quantitatively assessed, and the information of the flexibility potential $\mathcal{F}$ is shared with the power grid. At \textit{Step 2}, called \textit{Flexibility Exploitation}, based on the flexibility potential $\mathcal{F}$, the grid operator will generate a feasible DR request $\mathcal{D}$ to exploit the energy flexibility.

\subsection{Flexibility Assessment}
For this two-step design framework, the main difficulty lies in the first step: quantitatively characterizing and computing the flexibility potential of the system. Our work mainly focuses on solving grid congestion, which is a challenging issue for the power grid in the Netherlands. Hence, the DSM design mainly aims at reducing energy consumption during critical hours. 

While there is no standard definition of energy flexibility, the flexibility potential of a system can be measured mainly based on three factors that describe the shift in energy use: magnitude, time duration, and cost \cite{junker2018characterizing,soren17}. In our problem setup, the HP can only operate in two modes: on and off, which implies the magnitude of the flexibility reduction at each time instant is constant, the capacity of energy flexibility is thus proportional to the time duration that HP remains off. Consequently, we measure the flexibility capacity of the HPTES system as the time duration during which the HP can remain off without violating system constraints. 

\begin{figure}
    \centering
    \includegraphics[width = 0.8\linewidth]{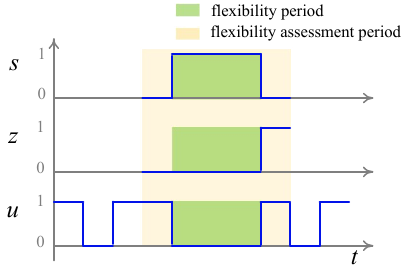}
    \caption{Schematic of the logic in \eqref{eq:flexi_cons} indicating flexibility periods.}
    \label{fig:logics}
\end{figure}

Here we denote $\mathcal{T}:=\{t_{f_1},\cdots,t_{f_n}\}$, called flexibility assessment period, as the set of consecutive time indices during which the flexibility of the HPTES system is assessed. Then, given a flexibility period, the time instants within $\mathcal{T}$ can be classified into three categories: the time instant before the flexibility period, at which the corresponding HP control input can be freely adjusted; the time instant within the flexibility period, during which the HP is off; and the time instant after the flexibility period, during which the HP control signal is free to be adjusted. To express the above logic and perform flexibility assessment, two sets of auxiliary decision variables are introduced: $\mathcal{S_T}:=\{s_t,t\in\mathcal{T}\}$ and $\mathcal{Z_T}:=\{z_t,t\in\mathcal{T}\}$. where $s_t\in\mathbb{B}$ and $z_t\in\mathbb{B}$ are binary decision variables. Based on $s_t$ and $z_t$, the following logic follows
\begin{itemize}
    \item $s_t = 0$ and $z_t = 0$: the time instant $t$ is before the flexibility period, and $u_t\in\mathbb{B}$.
    \item $s_t = 1$ and $z_t = 0$: the time instant $t$ is within the flexibility period, and $u_t = 0$.
    \item $s_t = 0$ and $z_t = 1$: the time instant $t$ is after the flexibility period, and $u_t\in\mathbb{B}$.
\end{itemize}
The above logic is illustrated by a diagram in Fig. \ref{fig:logics}, and can be expressed as the following mixed-integer linear constraints
\begin{subequations}\label{eq:flexi_cons}
    \begin{align}
        &u_t \leq 1 - s_t, \\
        &s_{t+1} \geq s_t - z_{t+1}, \\
        &s_t + z_t \leq 1,\\
        &z_{t+1} \geq z_t.
    \end{align}
\end{subequations}
Based on the definition of flexibility, the flexibility capacity is proportional to $\sum_{t\in\mathcal{T}}s_t$. Namely, the longer the period that the HP can remain off without violating system constraints, the larger the flexibility capacity. Combining the flexibility assessment with the economic MPC problem \eqref{eq:economic_mpc} leads to the following flexibility assessment formulation
\begin{subequations}\label{eq:flexi_comp}
    \begin{align}
       \min_{ u_t,s_k,z_k}\ & J_o - J_f \label{eq:obj}\\
        \text{s.t. } & x_{t+1} = f(x_{t},u_{t},d_{t}), \label{eq:sys_dyn}\\
        & x_t \text{ satisfy } \eqref{eq:temp_cons},\ u_t \text{ satisfy }\eqref{eq:switch},\\
        & \text{for  } t = 0,1,2,\cdots, N-1,\\
        & (u_k,s_k,z_k)\text{ satisfy }\eqref{eq:flexi_cons}, \forall k\in\mathcal{T}
    \end{align}
\end{subequations}
where $J_f := \lambda \sum_{t\in\mathcal{T}}s_t$ is the revenue of exploiting the energy flexibility for DSM services, and $\lambda > 0$ is a weighting parameter. By solving this optimization problem, the optimal flexibility period, which is defined as $\mathcal{F}:=\{t|s_t = 1\}$, can be computed. This information will be subsequently used in the \textit{Flexibility Exploitation} stage.
{\revise The optimization problem \eqref{eq:flexi_comp} computes the optimal flexibility commitment by balancing the profits of flexibility provision $J_f$ and the incurred operational cost $J_o$. Once $J_o$ is omitted, the optimal solution of this optimization problem quantifies the largest flexibility capacity of the system regardless of the incurred increase of operational cost. It should be pointed out that in this work the revenue of flexibility $J_f$ is defined to be proportional to the duration of flexibility, in order to demonstrate the effectiveness of the proposed two-step DSM framework. The objective function in \eqref{eq:obj} can be modified to consider different revenue formulations depending on different market choices and contractual options, such as maximal demand contract and interruptible load contract.

}

\subsection{Flexibility Exploitation}
After finishing the flexibility assessment, the information of the flexibility capacity $\mathcal{F}:=\{t|s_t=1\}$ will be shared with the power grid. Based on $\mathcal{F}$, a feasible DR request $\mathcal{R} = \{t|t\in\mathcal{F}\}\subseteq\mathcal{F}$ will be generated by the grid operator to specify the time instants that the HP needs to be off to provide DSM services. Namely,
\begin{equation}\label{eq:dr_request}
    u_t = 0,\quad\forall t\in\mathcal{R}.
\end{equation}
To achieve this DR request, a classical economic MPC problem with extra constraints \eqref{eq:dr_request} can be formulated to compute the HP control signals. Since the DR request $\mathcal{D}$ is feasible w.r.t. the flexibility capacity $\mathcal{F}$, there is always an admissible control input sequence, i.e., a solution of the optimization problem \eqref{eq:flexi_comp}, to meet the DR request without violating system constraints.  

\remarknew{ By solving \eqref{eq:flexi_comp}, the optimal flexibility period, which is defined as $\mathcal{F}:=\{t|s_t=1\}$, will be computed. If the operational cost $J_o$ is omitted from the objective function, the largest flexibility capacity of the system, i.e., the longest period during which the HP can remain off without violating constraints, is obtained. It should be highlighted that the formulation of flexibility assessment in \eqref{eq:flexi_comp} is independent of the system dynamics as well as the economic MPC formulation, which means that the proposed DSM design framework is feasible to any economic MPC design by only considering extra linear constraints \eqref{eq:flexi_cons}. Hence, this design framework can be easily incorporated into other MPC formulations without inducing high design complexity and computational burden.}
\section{Numerical and Experimental Results}
\begin{figure}[tb]
    \centering
\begin{subfigure}[b]{0.45\linewidth}
    \centering
        \includegraphics[height=0.7\linewidth]{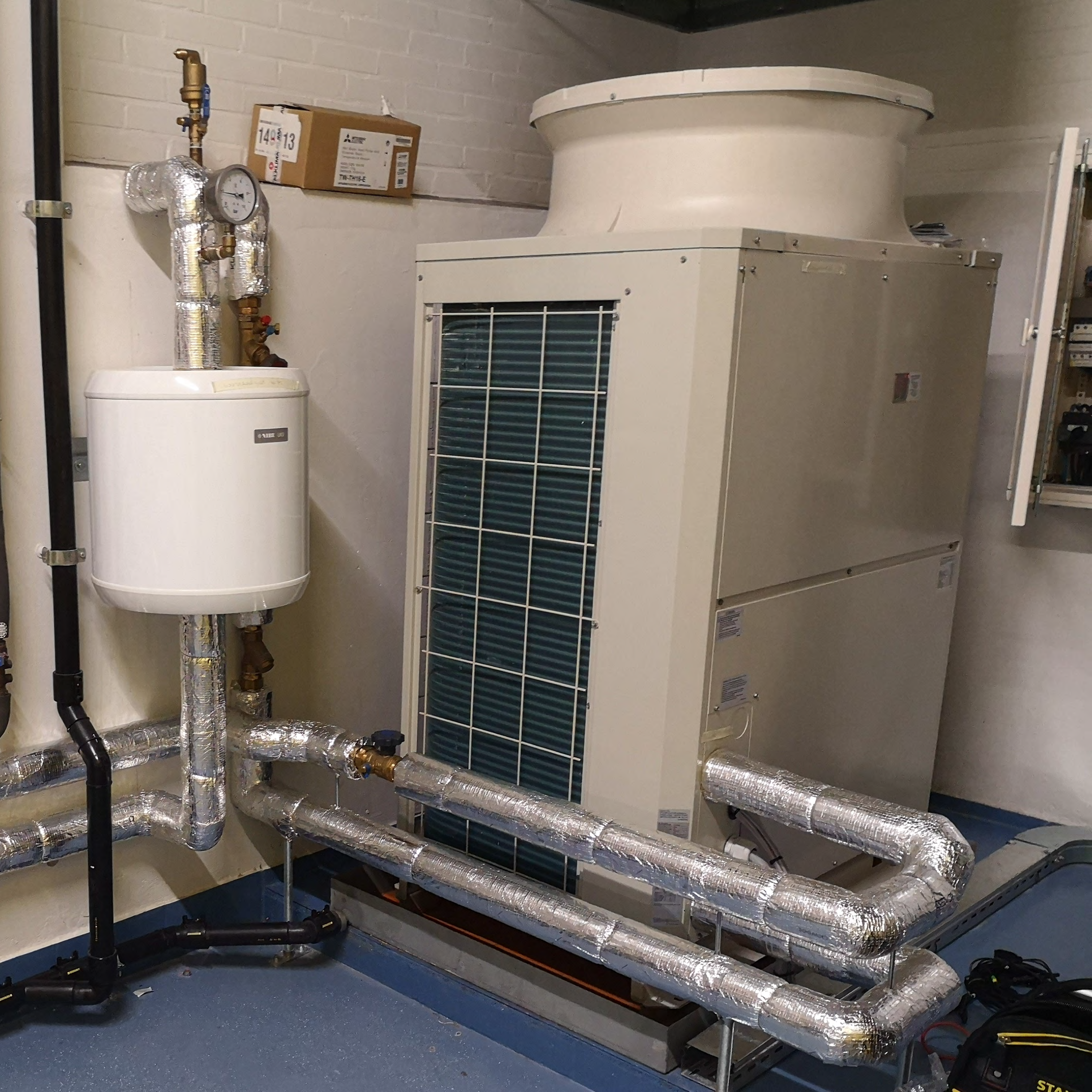}
        \caption{Air-to-water HP.}
        \label{fig:real2}
    \end{subfigure}\quad
    \begin{subfigure}[b]{0.45\linewidth}
    \centering
        \includegraphics[height=0.7\linewidth]{real_system_1.pdf}
        \caption{Two 500L water tanks.}
        \label{fig:real1}
    \end{subfigure}
    \caption{The HPTES system used in our numerical and experimental case studies.}
    \label{fig:real}
\end{figure}

In this section, the modeling approaches as well as the MPC-based DSM strategy will be numerically and experimentally tested. The investigated HPTES system is based on a real installation, as shown in Fig. \ref{fig:real}, that is designed to provide domestic hot water for fitness rooms and dining rooms in an office building in the Netherlands.  This HPTES system consists of two 500-liter water tanks and an air-to-water heat pump. The HP is only operated during working hours from 7:00 to 17:30 on weekdays.

The numerical and experimental results are presented in 4 subsections including the control-oriented modeling results of the HPTES system, simulation \& experiment settings and MPC configurations, simulation results of the proposed MPC schemes, and real-world experiments of the proposed MPC schemes. All MPC problems for simulations and experiments are modeled using the Python package {\tt Pyomo} \cite{hart2011pyomo} and are solved using Gurobi 10.0.1 \cite{gurobi} on a 2.3 GHz 8-Core Intel Core i9 CPU with 32 GB 2667 MHz DDR4 RAM.

\subsection{Control-Oriented Models for HPTES system}
\subsubsection{COP Model for HP}
\begin{figure}
    \centering
    \includegraphics[width=0.8\linewidth]{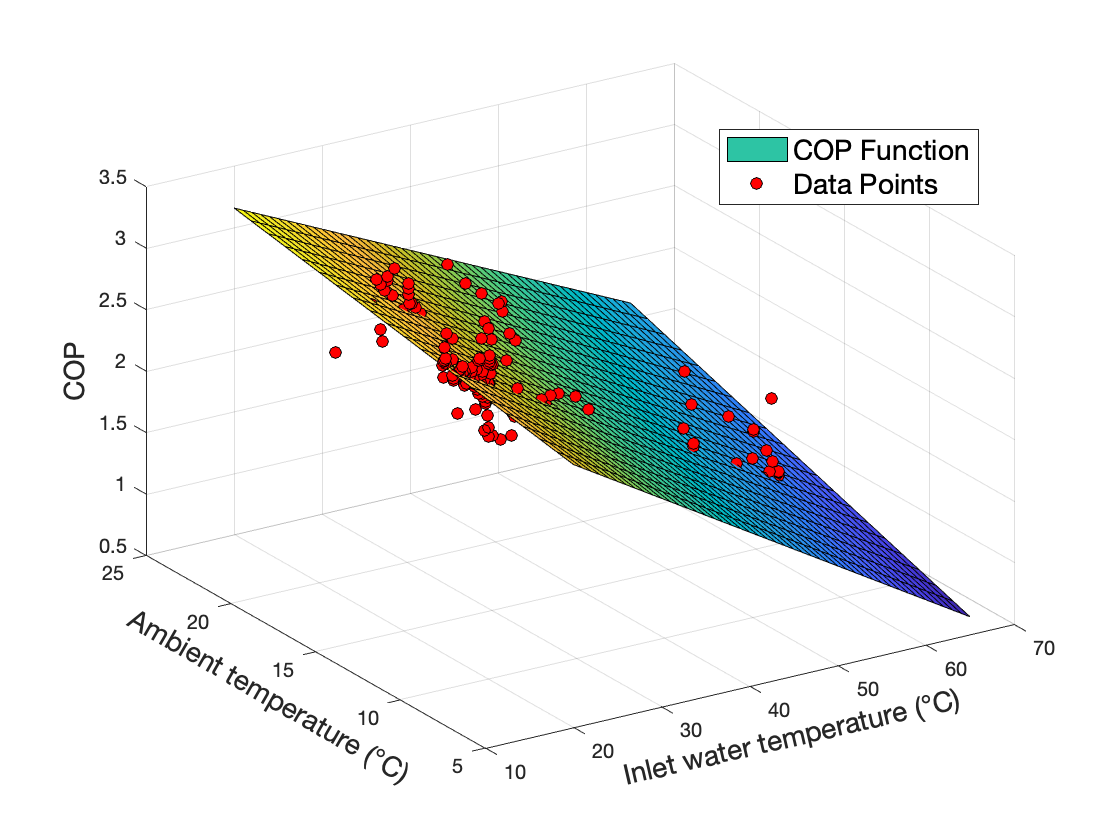}
    \caption{Approximated COP function in \eqref{eq:cop2} and measured data samples.}
    \label{fig:cop}
\end{figure}

The COP model is trained using 176 data points to minimize the sum of the square of prediction errors
\begin{equation}
    \min_{a_1,a_2,a_3,a_4} \sum_{i}(f_{\text{COP}} - {\text{COP}}_i)^2
\end{equation}
where $f_{\text{COP}}$ is the COP function defined in \eqref{eq:cop2} and $\text{COP}_i$ is the $i$-th COP data sample. The optimal COP parameters are $a_1 = 3.3297, a_2 = -0.0423, a_3 =  0.0219, a_4 = 0.0003$, which lead to root mean square error (RMSE) 0.21 and variance accounted for (VAF) $83.93\%$.  Fig. \ref{fig:cop} presents the computed COP function and the training data samples, from which it can be seen that the bilinear COP model \eqref{eq:cop2} can properly approximate the non-constant property of the HP COP.

\subsubsection{Prediction of Hot Water Consumption}
\begin{figure}
    \centering
    \includegraphics[width=0.9\linewidth]{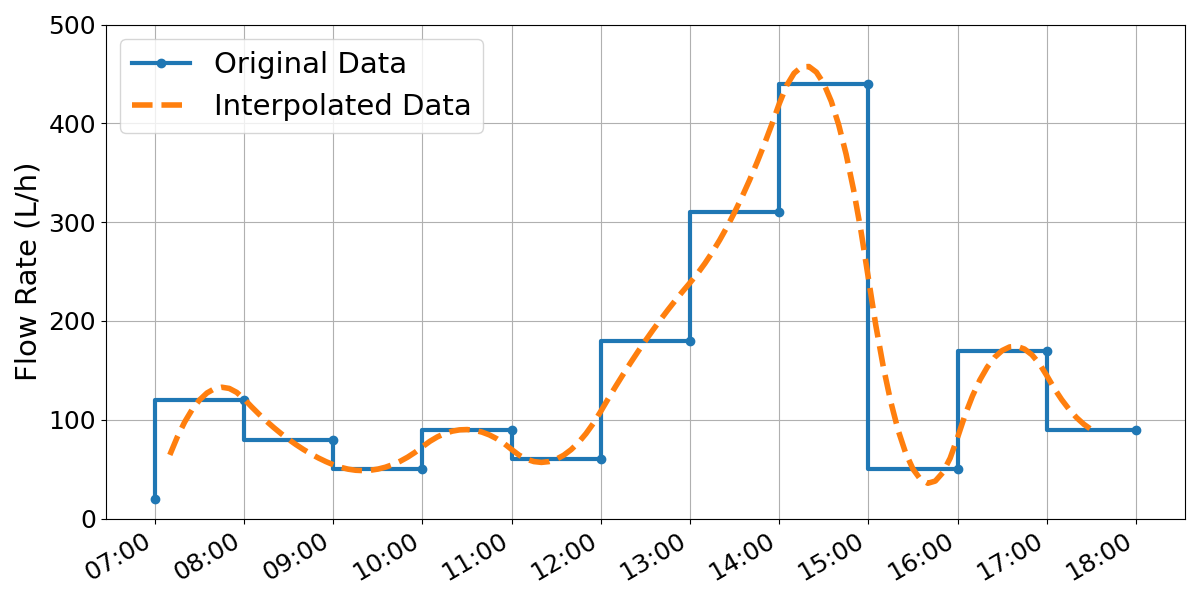}
    \caption{Interpolation of the hourly water consumption.}
    \label{fig:interp}
\end{figure}

For our investigated real HPTES system, due to the limitation of the existing sensors installment, the hot water consumption is only measured at an hourly level. However, the water temperature measurements as well as our control implementation are performed at a minute level. This discrepancy necessitates the transformation of the resolution of the hot water consumption measurement from hourly level to minute level. In this work, polynomial interpolation is performed for the hourly water consumption measurements to compute the corresponding minute-level water usage approximations. Specifically, the polynomial function from the Python package {\tt Pandas} is exploited \cite{pandas_api}. Fig. \ref{fig:interp} shows the real hourly water usage measurements and the corresponding minute-level water usage approximations.

After computing the minute-level water consumption, the SARIMA model is further implemented to predict the hot water usage. Given the inherent seasonality in the hot water consumption patterns, we combine two SARIMA models with different seasonalities to generate the prediction of hot water consumption: 
\textit{1)} $Y_{\text{daily}}$, with a seasonality of 24 hours, to capture the daily patterns;
\textit{2)} $Y_{\text{weekly}}$, with a seasonality of 1 week, to capture possible weekly water consumption patterns. 
The combined model is expressed as:
\begin{equation}\label{eq:sarima}
Y_{\text{combine}} = \alpha Y_{\text{weekly}} + (1-\alpha) Y_{\text{daily}}
\end{equation}
where $\alpha\in(0,1)$ is a weighting parameter, and $Y_{\text{combine}}$ is the combined prediction of hot water consumption. The optimal parameters for each SARIMA model are computed using constrained maximum likelihood estimation by assuming the forecast error follows a Gaussian distribution.

\begin{figure}[h]
    \centering
   \includegraphics[width=0.95\linewidth]{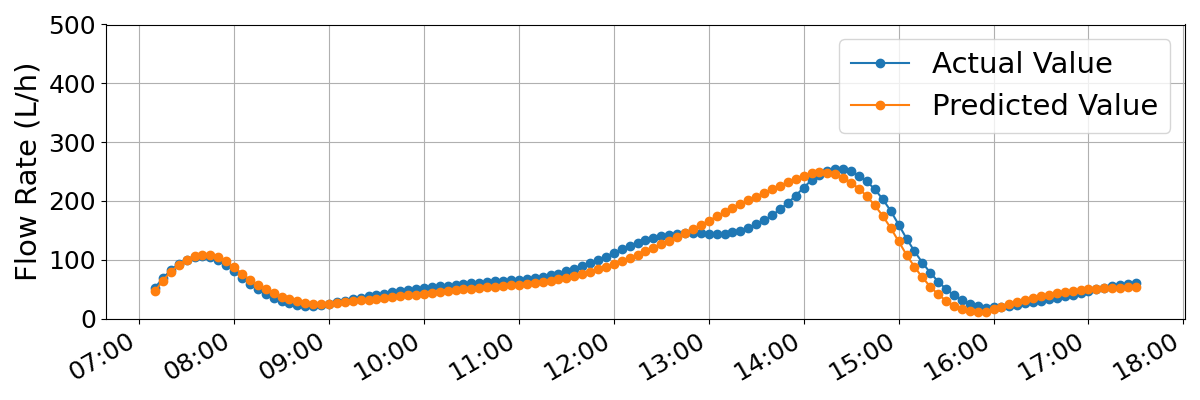}
    \caption{Hot water consumption and its prediction.}
    \label{fig:demand}
\end{figure}

Fig. \ref{fig:demand} shows the real hot water consumption measurement and the prediction of hot water consumption via \eqref{eq:sarima} at a specific day. It can be seen that our prediction model can properly capture the water consumption patterns and achieve an acceptable prediction error for real implementation. In the upcoming sections, the predicted hot water consumption will be used for MPC design. It will be shown that the proposed MPC approach works well even in the presence of prediction errors for hot water consumption.

\subsubsection{Stratified Water Layer Model for Water Tanks}
\begin{figure}
            \centering
            \includegraphics[width=\linewidth]{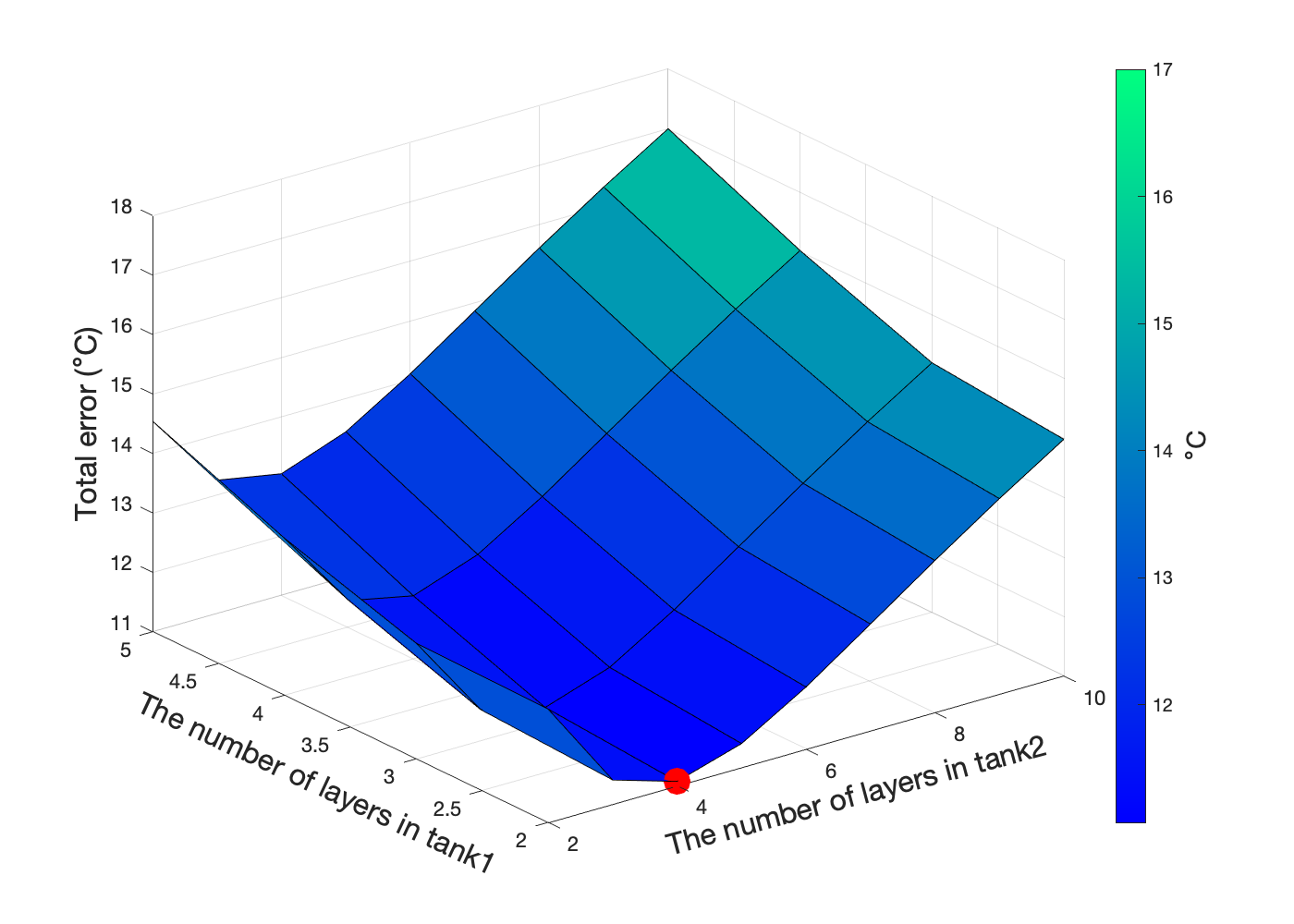}
            \caption{Total RMSE of predicted temperatures as a function of the different numbers of layers in the hot water storage tank models. }
            \label{fig:layer2}
        \end{figure}

For the control-oriented model of the thermal tank in \eqref{eq:tank_model}, the model parameters to be identified are $(m_j,R_{j,j+1},R_{j-1,j},R_{w,j})$ and the number of the stratified layers. These model parameters are identified by optimizing the root mean square of the prediction error of the water temperature for the top and bottom layers of Tank 1 and the bottom layer of Tank 2, which are available for measurement in our system. The optimal numbers of stratified water layers for two water tanks are computed by grid search. Fig. \ref{fig:layer2} gives the total RMSE of the temperature approximation error w.r.t. different combinations of the water layers for two water tanks, which identifies (2,4) as the best combination. The detailed models and corresponding parameter values are presented in our previous work \cite{10666588}.

It should be highlighted that the results in Fig. \ref{fig:layer2} are counter-intuitive since the prediction error is expected to decrease as the number of layers is increased for the two tanks. A possible reason is that the approximated hot water consumption, which is used for developing the water tank models, computed from hourly measurements might not reflect the actual hot water consumption at minute level. This discrepancy can then influence the real-time thermal dynamics of water tanks, and consequently the prediction error of the water temperature.

\begin{figure}
    \centering
    \includegraphics[width=1\linewidth]{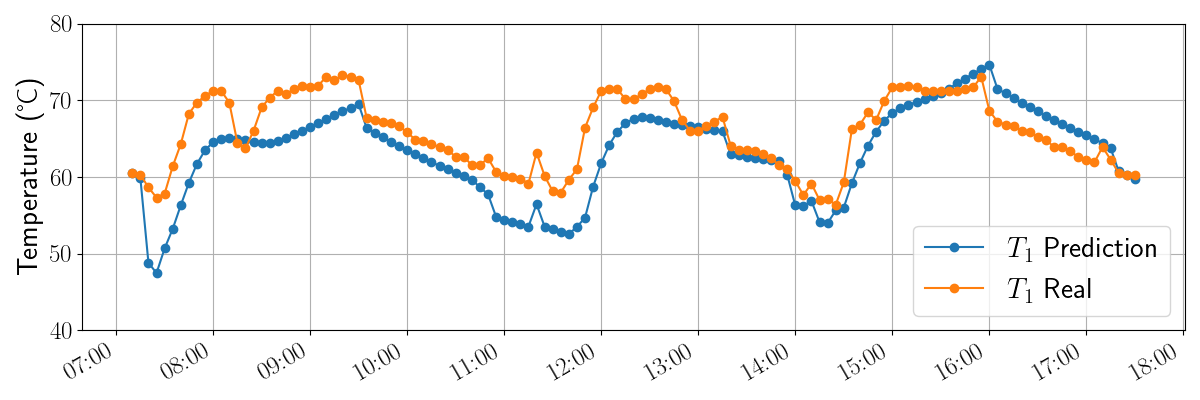}
    \caption{Water temperature for the top layer of Tank 1 $T_1$, and its prediction.}
    \label{fig:T1}
\end{figure}

Fig. \ref{fig:T1} shows the real measurements of the water temperature for the top layer of Tank 1, which is crucial in our system in determining the temperature of the supplied hot water. It can be seen that our prediction model can reflect the trend of the temperature dynamics very well despite the lack of an accurate measurement of hot water consumption and sufficient temperature sensors, which could be common issues in practical HPTES installations. In the next section, it will be shown that this approximate prediction model still works satisfactorily for subsequent MPC design.

\subsection{Simulation \& Experiment Settings and MPC Configurations}
\begin{figure}
    \centering
    \includegraphics[width=\linewidth]{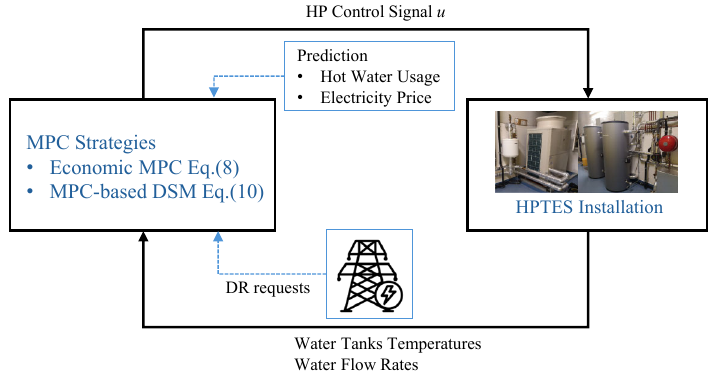}
    \caption{\revise Simulation and experiment diagram.}
    \label{fig:simu_diagram}
\end{figure}

{\revise Fig. \ref{fig:simu_diagram} depicts the diagram for implementing our simulation and experiment of the proposed MPC strategies. 
At the beginning of the simulation or implementation, electricity price signals and hot water consumption predictions are updated. Then, the MPC solution is computed every 5 minutes. At each sampling instant, the real water tank temperature and water flow rates are used to update the control-oriented model. For each MPC update, if a new DSM request is detected and no existing DR request is being served, the DSM scheme will be activated to meet the requirement of the power grid. Otherwise, the economic MPC scheme will control the HP to optimize system operation.}

\begin{figure}[h]
    \centering
   \includegraphics[width=\linewidth]{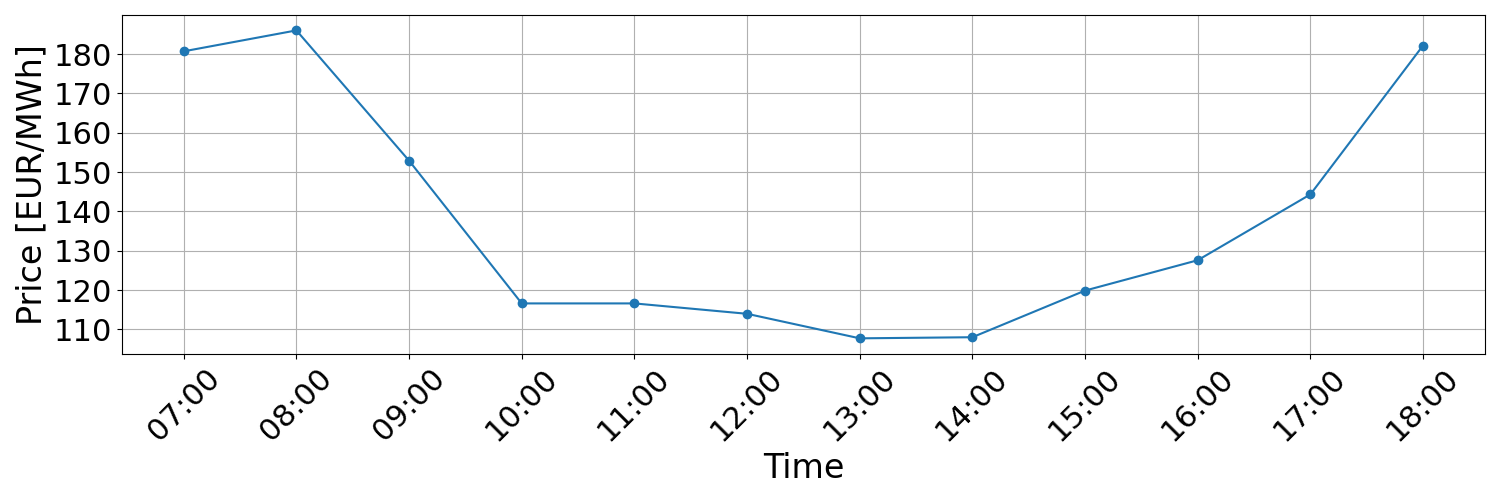}
    \caption{Electricity price used in tests.}
    \label{fig:price}
\end{figure}

In our system, the temperature of the hot water supplied to the building is required to be within $55^\circ$C - $75^\circ$C to mitigate the risk of legionella bacteria and to ensure system safety. Besides, it is preferred to keep the temperature above $60^\circ$C for most of the operational time to provide qualified hot water. For the real-time implementation of MPC, it is essential to ensure that the MPC problems in \eqref{eq:economic_mpc} and \eqref{eq:flexi_comp} are always feasible. Due to the existence of uncertainties, the hard constraints \eqref{eq:temp_cons} and \eqref{eq:switch} might result in infeasible optimization problems. To address this issue, we alternatively consider the following soft constraints
\begin{subequations}
    \begin{align}
        &55 - \delta_1 \leq T_1 \leq 75 + \delta_1 \\
        &T_1 \geq 60 - \delta_2
    \end{align}
\end{subequations}
where $\delta_1> 0$ and $\delta_2> 0$ are slack variables to relax the hard constraints and are penalized in the objective function $J_o$. The operational cost function $J_o$ then becomes 
\begin{equation}
    J_o := \sum_{t=1}^Ne_tP_ru_t + M_1\delta_1 + M_2\delta_2
\end{equation}
where $e_t$ is the electricity price, $M_1\geq 0$ and $M_2\geq 0$ are large constants to penalize the slack variables $(\delta_1,\delta_2)$. The electricity price used in our simulation is depicted in Fig. \ref{fig:price}, which is a typical electricity price pattern in the Dutch electricity market.

{\revise For our proposed MPC strategies, the prediction horizon is selected as $6$ hours for the economic MPC scheme \eqref{eq:economic_mpc}, and 4 hours for flexibility assessment \eqref{eq:flexi_comp}. This choice is adopted to balance the computational costs and control performance, considering the fact that the HP is operated for about 10 hours per day.
In addition, practical DR request periods for solving peak hour grid congestion last generally for less than an hour, and the real-time control update period of our MPC schemes is 5 minutes.} 

One practical challenge in implementing MPC is ensuring that the corresponding optimization problems can be solved efficiently for real-time implementation. Good control performance of MPC generally requires considering a relatively long prediction horizon, which will increase the computational burden. To balance the computational efficiency and control performance, the \textit{move blocking strategy} is utilized in our MPC implementation. Unlike the conventional MPC design, where fixed time steps are applied within the whole prediction horizon, the \textit{moving blocking strategy} involves segmenting the prediction horizon into time windows with different lengths of time steps \cite{schwickart2016flexible}. For the blocks that are closer to the current time instant, the corresponding time step is smaller. This strategy is based on the premise that the necessity for precise prediction and control diminishes for further predictions in the future since only the first MPC input signal is applied and the control signal at the next time step is recomputed with updated information.

\begin{figure}[h]
    \centering
   \includegraphics[width=\linewidth]{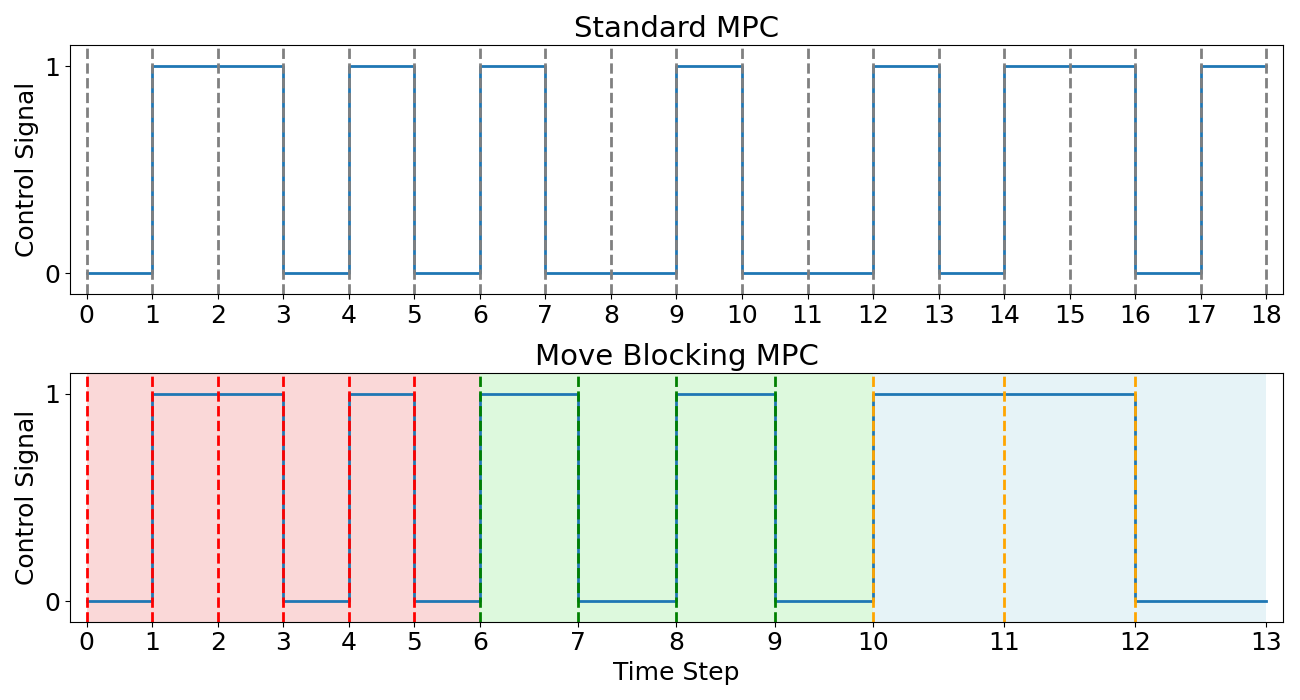}
    \caption{Standard MPC and move blocking MPC in a 6-hour prediction horizon.}
    \label{fig:moveblocking}
\end{figure}

Fig. \ref{fig:moveblocking} gives a graphical illustration of the standard MPC and \textit{move blocking} MPC. It can be seen that the standard MPC keeps a fixed time step within the prediction horizon. In contrast, \textit{move blocking} MPC has varying-sized time steps. With the same prediction horizon length, \textit{moving blocking} MPC entails fewer decision variables than the standard MPC, and hence can consider a longer prediction horizon with reduced computational burden for solving the corresponding MPC problem. As shown in Fig. \ref{fig:moveblocking}, in our simulations and experiments, the prediction horizon for the economic MPC formulation \eqref{eq:economic_mpc} is 6 hours with 3 blocks. The first block has 6 time steps with each time step 20 mins. The second and the third blocks have 4 and 3 steps with corresponding time steps as 30 mins and 40 mins, respectively. As shown in Table \ref{tab:mb_time}, with the move-blocking strategy, the average computational time for solving the economic MPC problem \eqref{eq:economic_mpc} over a simulation day is reduced from 72.96s to 25.77s, highlighting the improved computational efficiency.

\begin{table}[htb]
    \centering
    \caption{Computation time of standard MPC and \text{move blocking} MPC.}
    \resizebox{\linewidth}{!}{
    \begin{tabular}{lcc}\toprule\hline
         &  Standard MPC & Move Blocking MPC\\\hline
        Minimal computation time (s) & 11.74& 10.81\\
        Maximal computation time (s) & 301.19& 56.58\\
          Average computation time (s) & 72.96 & 25.77\\\hline
    \end{tabular}}
    \label{tab:mb_time}
\end{table}


\subsection{Simulation Results of MPC Schemes}
\subsubsection{Simulation Results of MPC for Economic Operation} 
For the proposed MPC strategies, the performance of the economic MPC formulation \eqref{eq:economic_mpc} without considering DSM is tested first. To show the efficacy of the proposed MPC approach, the results of the \textit{Rule-Based Approach} that is currently used to operate the real HPTES system are also presented. With the \textit{Rule-Based Approach}, the HP is turned on if the water temperature at the top layer of Tank 1 is lower than 62$^\circ$C and is turned off until the water temperature at the bottom layer of Tank 2 is higher than 62$^\circ$C.

\begin{figure}[tb]
    \centering
   \includegraphics[width=1\linewidth]{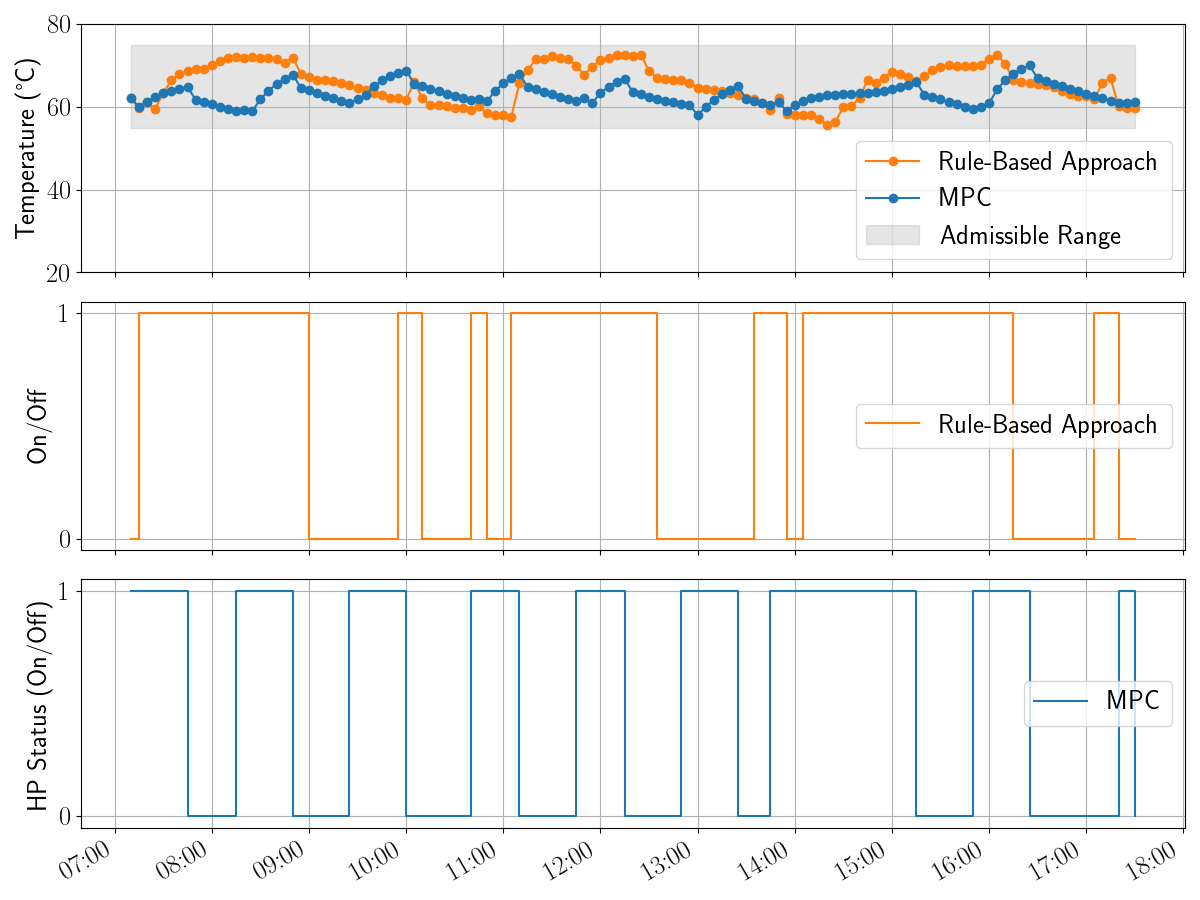}
    \caption{Simulation results for Economic Operation: (top) water temperature at the top layer of Tank 1, (middle) HP status with rule-based scheme, (bottom) HP status with MPC.}
    \label{fig:simu_Econ}
\end{figure}

The real water consumption and its prediction utilized in our simulation is depicted in Fig. \ref{fig:demand}. The computational time for solving the MPC problem is given in Table \ref{tab:simu_Econ}. Simulation results are shown in Fig. \ref{fig:simu_Econ} and are summarized in Table \ref{tab:simu_Econ}. From Fig. \ref{fig:simu_Econ}, it is clear that the MPC approach can maintain the supply water temperature within the admissible range $[55^\circ\text{C},75^\circ\text{C}]$. It can be seen from Table \ref{tab:simu_Econ} that compared with the rule-based approach, the energy cost of MPC is reduced by about 14\% and the energy consumption by $13\%$ while also leading to less constraint violations (defined as the violation of the desired lower bound of water temperature $60^\circ$C).

\begin{table}[htb]
    \centering
    \caption{Simulation results of MPC and rule-based schemes for Economic Operation.}
    \resizebox{\linewidth}{!}{
    \begin{tabular}{lcc}\toprule\hline
         &  Rule-Based Approach & MPC\\\hline
        Average Water Temperature ($^{\circ}$C) & 65.53& 63.11\\
        Maximal Constraint Violation ($^{\circ}$C) & 4.43& 1.94\\
         Energy Consumption (kWh) & 64.17 (100\%) & 55.83 (87.00\%)\\
          Energy Cost (Euro) & 8.69 (100\%) & 7.45 (85.73\%)\\\hline
    \end{tabular}}
    \label{tab:simu_Econ}
\end{table}
\subsubsection{Simulation Results of MPC for DSM} Based on the economic MPC formulation, we augment the design to further consider DSM. The data for electricity price and water consumption used are the same as for the economic MPC implementation. For the flexibility assessment problem \eqref{eq:flexi_comp}, the prediction horizon is selected as 4 hours, and the first 3 hours are defined as the flexibility assessment period. This time interval is chosen to ensure that the flexibility potential of the system can be fully explored while leaving sufficient time to prepare for the next flexibility task. In our simulation and experiments, synthetic DR requests are generated from the power grid every 3 hours so that the flexibility assessment and exploitation tasks are performed (3 tasks in total per day). The objective function for \eqref{eq:flexi_comp} is selected as $J_f=\sum_{t\in\mathcal{T}}s_t$ without considering the operational cost $J_o$, which implies that the largest flexibility potential of the HPTES system, i.e., the longest time period that the heat pump can be off without violating system constraints, is computed. For flexibility exploitation, virtual DR requests are set as $\mathcal{D} = \mathcal{F}$, which request the largest possible energy reduction during the flexibility period. This extreme scenario is selected for testing the robustness of our proposed approach. 

\begin{figure}[tb]
    \centering
   \includegraphics[width=1\linewidth]{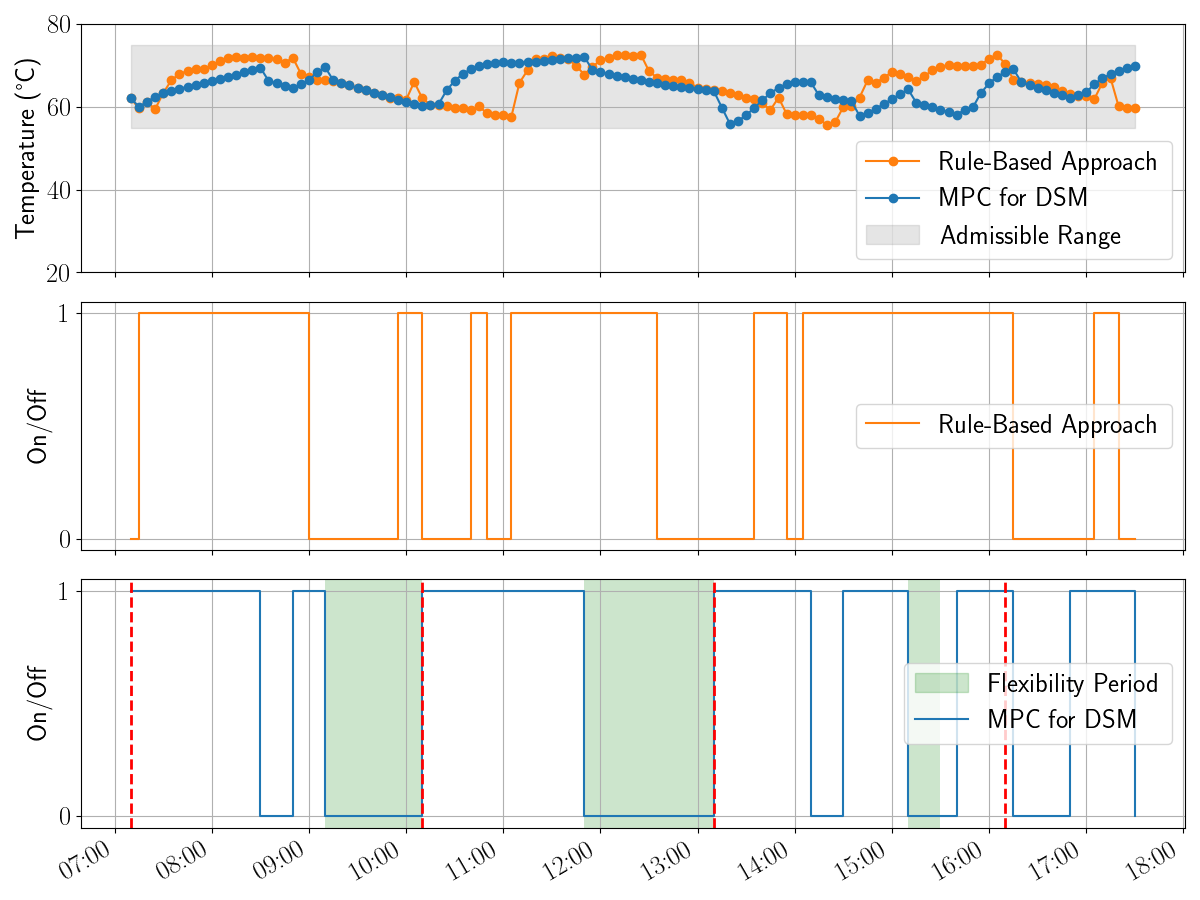}
    \caption{Simulation results for DSM: (top) water temperature at the top layer of Tank 1, (middle) HP status with rule-based scheme, (bottom) HP status with MPC.}
    \label{fig:simu_DSM}
\end{figure}

Simulation results for DSM are 
shown in Fig. \ref{fig:simu_DSM} and Table \ref{tab:simu_DSM}. In Fig. \ref{fig:simu_DSM}, the starting and ending times of flexibility assessment are indicated via dashed red lines, and the flexibility periods as green shaded areas. Within all flexibility periods, the HP is required to remain off to meet the DR requests. It is clear that all DR requests are satisfied while respecting system constraints. Furthermore, as shown in Table \ref{tab:simu_DSM}, compared with the rule-based control approach, which is incapable of implementing DSM, our proposed MPC approach can not only achieve DSM but also leads to less energy consumption. This result highlights the benefits of the proposed MPC-based approach in cost saving even with the provision of DSM services, which will force the operation of HP to deviate from its economically optimal pattern, and sacrifice economic operational performance.

\begin{table}[htb]
    \centering
    \caption{Simulation results of MPC and rule-based schemes for DSM.}
    \resizebox{\linewidth}{!}{
    \begin{tabular}{lcc}\toprule\hline
         &  Rule-Based Approach & MPC\\\hline
        Average Water Temperature ($^{\circ}$C) & 65.53& 64.85\\
        Maximal Constraint Violation ($^{\circ}$C) & 4.43& 4.12\\
         Energy Consumption (kWh) & 64.17 (100\%) & 62.50 (97.40\%)\\
          Energy Cost (Euro) & 8.69 (100\%) & 8.41 (96.78\%)\\\hline
    \end{tabular}}
    \label{tab:simu_DSM}
\end{table}

\subsection{Experimental Results of MPC Schemes}
\subsubsection{Experimental Results of MPC for Economic Operation}
\begin{figure}[tb]
    \centering
   \includegraphics[width=0.95\linewidth]{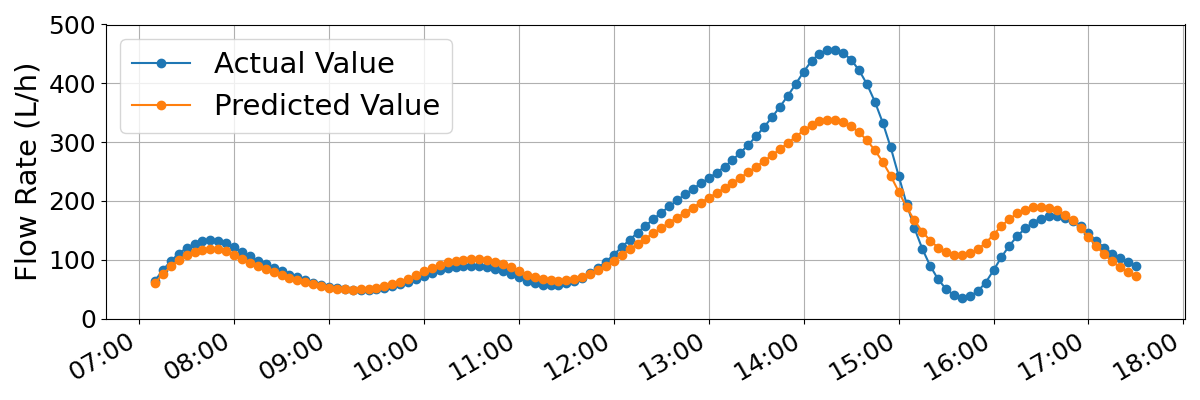}
    \caption{Hot water consumption and its prediction in the experiment for economic operation (MAE of prediction: 28.32 L/h).}
    \label{fig:imple_Econ_SARAMA}
\end{figure}

\begin{figure}[tb]
    \centering
   \includegraphics[width = 0.95\linewidth]{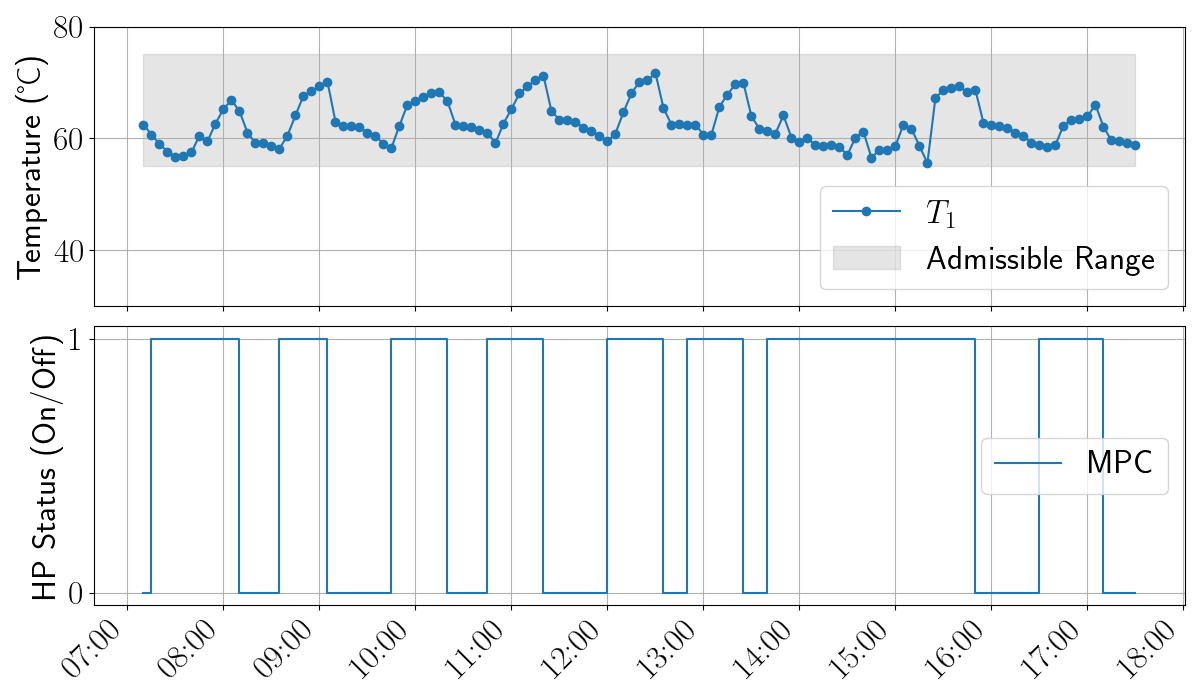}
    \caption{Experimental results of MPC for economic operation: (top) water temperature at the top layer of Tank 1; (bottom) HP status.}
    \label{fig:imple_Econ}
\end{figure}

In order to validate the practical viability of our proposed schemes, experiments are conducted in a real-world HPTES installation as shown in Fig. \ref{fig:real}. The real hot water consumption and its predictions during the experiment are shown in Fig. \ref{fig:imple_Econ_SARAMA}, from which we can conclude that the prediction reflects the real usage reasonably well but with non-negligible prediction error, especially during peak hours. Fig. \ref{fig:imple_Econ} depicts the water temperature of the supplied hot water and the HP control signals. It is clear that the proposed MPC approach successfully maintains the water temperature within the admissible range regardless of the non-negligible prediction error of water usage. In addition, based on our computation, the MPC approach results in an operational cost of 8.64 euros, which is lower than the 9.63 euros achieved by the existing rule-based approach with similar hot water consumption profile.

\subsubsection{Experimental Results of MPC for DSM}
\begin{figure}[htb]
    \centering
   \includegraphics[width=0.95\linewidth]{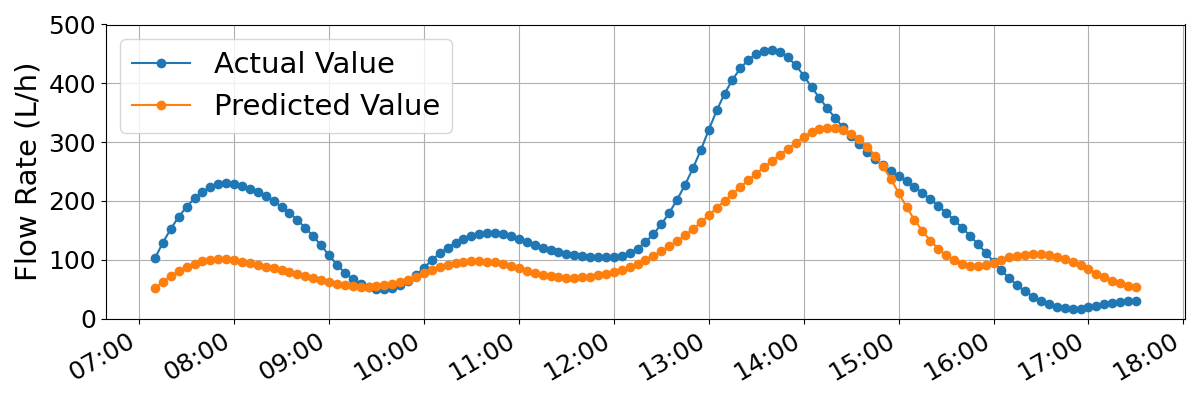}
    \caption{Hot water consumption and its prediction in the experiment for DSM (MAE of prediction: 64.93 L/h).}
    \label{fig:imple_DSM_SARAMA}
\end{figure}

This section presents the experimental results of our proposed approach for DSM in Section \ref{sec:mpc}.  Fig. \ref{fig:imple_DSM_SARAMA} depicts the real and predicted hot water consumption during the experiment day. For this specific day, an abnormal real hot water consumption happened. Specifically, it can be seen from Fig. \ref{fig:imple_DSM_SARAMA} that the peak hour started earlier than usual so that a relatively larger prediction error occurred compared to our previous case. While the prediction performance of water consumption is unsatisfactory, this scenario provided a good case to test the robustness of our proposed approach. The supplied hot water temperature and the HP control signals are plotted in Fig. \ref{fig:imple_DSM_SARAMA}. It can be seen that despite the large water usage prediction error, our proposed approach still performs well. The promised DR services were achieved while keeping the water temperature within the admissible range during almost the entire experiment. The only exception occurred around 14:00, at which the water temperature was slightly below the admissible lower bound due to the large prediction error of water usage.

\begin{figure}[htb]
    \centering
   \includegraphics[width = 0.95\linewidth]{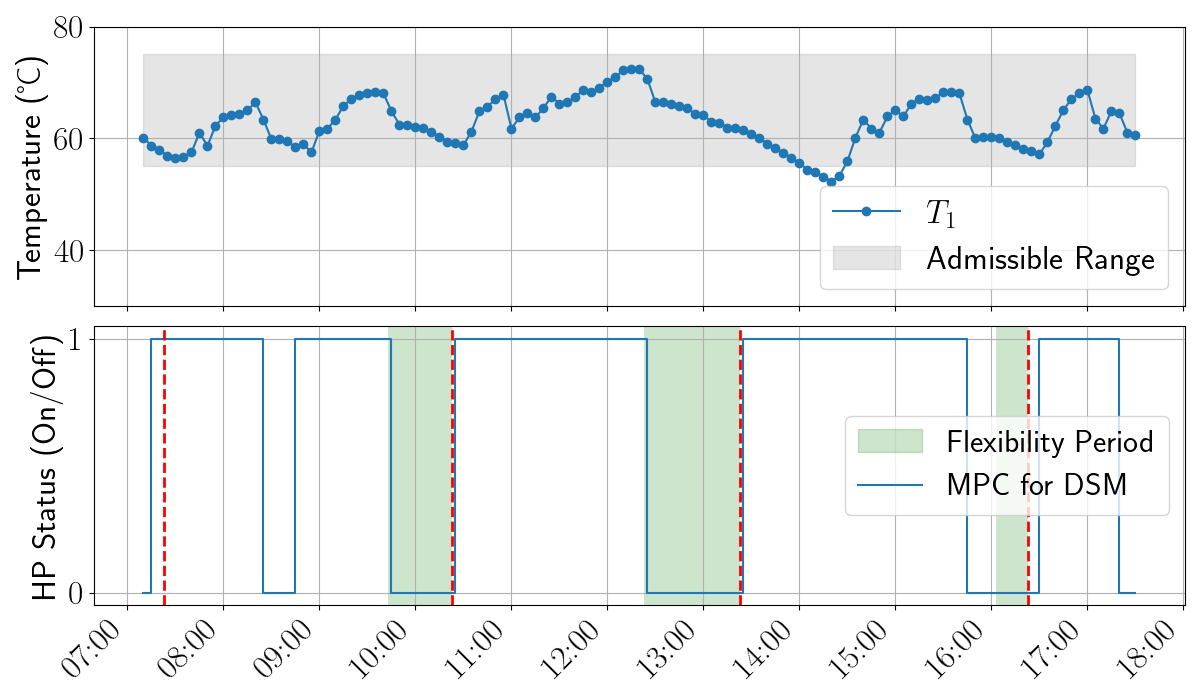}
    \caption{Experimental results of MPC for DSM: (top) water temperature at the top layer of Tank 1; (bottom) HP status.}
    \label{fig:imple_DSM}
\end{figure}

\section{Conclusions}\label{sec:conclusion}{\revise
This paper proposed an energy-flexible MPC design framework that leverages the energy flexibility of HPTES systems for DSM, and provided detailed procedures for control-oriented modeling of the main components of the system. The introduced control-oriented models are computationally tractable for MPC design and are adaptable to accommodate different HPTES configurations. Notably, our proposed MPC strategy can quantitatively assess the energy flexibility of HPTES systems by solving a typical economic MPC problem with additional mixed-integer linear constraints, ensuring that any admissible DR requests can be achieved while respecting system constraints. The proposed approach offers a systematic framework for modeling and operating HPTES systems using an advanced control strategy to achieve economic operation and DSM. The efficacy of our proposed approach is validated through both numerical and experimental results based on a real-world HPTES installation. The results demonstrate that the proposed MPC schemes are effective in achieving economical operation and energy-flexible DSM. Moreover, they confirm the practical viability of the future-proof S2 standard, enabling versatile and privacy-secured exploitation of energy flexibility in the built environment.
}

\section{Acknowledgement}
The authors would like to thank Joep van der Velden, Kevin de Bont, and Jan-Willem Dubbeldam from Kropman B.V. and Jos Ruijter from a.s.r. Nederland for their support with the implementation.

\bibliographystyle{IEEEtran.bst}
\bibliography{ref}

\begin{thebibliography}{10}
\providecommand{\url}[1]{#1}
\csname url@samestyle\endcsname
\providecommand{\newblock}{\relax}
\providecommand{\bibinfo}[2]{#2}
\providecommand{\BIBentrySTDinterwordspacing}{\spaceskip=0pt\relax}
\providecommand{\BIBentryALTinterwordstretchfactor}{4}
\providecommand{\BIBentryALTinterwordspacing}{\spaceskip=\fontdimen2\font plus
\BIBentryALTinterwordstretchfactor\fontdimen3\font minus \fontdimen4\font\relax}
\providecommand{\BIBforeignlanguage}[2]{{%
\expandafter\ifx\csname l@#1\endcsname\relax
\typeout{** WARNING: IEEEtran.bst: No hyphenation pattern has been}%
\typeout{** loaded for the language `#1'. Using the pattern for}%
\typeout{** the default language instead.}%
\else
\language=\csname l@#1\endcsname
\fi
#2}}
\providecommand{\BIBdecl}{\relax}
\BIBdecl

\bibitem{dutchgov2022}
\BIBentryALTinterwordspacing
{Dutch Government}, ``Klimaatnota 2022,'' 2022. [Online]. Available: \url{https://www.rijksoverheid.nl/documenten/publicaties/2022/11/01/klimaatnota-2022}
\BIBentrySTDinterwordspacing

\bibitem{bunning2022robust}
F.~B{\"u}nning, J.~Warrington, P.~Heer, R.~S. Smith, and J.~Lygeros, ``Robust {MPC} with data-driven demand forecasting for frequency regulation with heat pumps,'' \emph{Control Engineering Practice}, vol. 122, p. 105101, 2022.

\bibitem{Abl:45}
\BIBentryALTinterwordspacing
{European Commission, Directorate General for Energy}, ``Clean energy for all {Europeans} package,'' 2019. [Online]. Available: \url{https://energy.ec.europa.eu/topics/energy-strategy/clean-energy-all-europeans-package_en}
\BIBentrySTDinterwordspacing

\bibitem{EU2009}
\BIBentryALTinterwordspacing
{European Parliament and Council of the European Union}, ``{Directive 2009/28/EC of the European Parliament and of the Council of 23 April 2009 on the promotion of the use of energy from renewable sources and amending and subsequently repealing Directives 2001/77/EC and 2003/30/EC},'' 2009. [Online]. Available: \url{https://energy.ec.europa.eu/topics/renewable-energy/renewable-energy-directive-targets-and-rules/renewable-energy-directive_en}
\BIBentrySTDinterwordspacing

\bibitem{hp2022}
\BIBentryALTinterwordspacing
{European Commission}, ``Heat pumps are key to enabling the clean energy transition and achieving the {EU}’s carbon neutrality goal by 2050,'' 2022. [Online]. Available: \url{https://energy.ec.europa.eu/topics/energy-efficiency/heat-pumps_en}
\BIBentrySTDinterwordspacing

\bibitem{ermel2022thermal}
C.~Ermel, M.~V. Bianchi, A.~P. Cardoso, and P.~S. Schneider, ``Thermal storage integrated into air-source heat pumps to leverage building electrification: A systematic literature review,'' \emph{Applied Thermal Engineering}, p. 118975, 2022.

\bibitem{kuboth2019experimental}
S.~Kuboth, F.~Heberle, T.~Weith, M.~Welzl, A.~K{\"o}nig-Haagen, and D.~Br{\"u}ggemann, ``Experimental short-term investigation of model predictive heat pump control in residential buildings,'' \emph{Energy and Buildings}, vol. 204, p. 109444, 2019.

\bibitem{d2019model}
F.~D'Ettorre, P.~Conti, E.~Schito, and D.~Testi, ``Model predictive control of a hybrid heat pump system and impact of the prediction horizon on cost-saving potential and optimal storage capacity,'' \emph{Applied Thermal Engineering}, vol. 148, pp. 524--535, 2019.

\bibitem{kajgaard2013model}
M.~U. Kajgaard, J.~Mogensen, A.~Wittendorff, A.~T. Veress, and B.~Biegel, ``Model predictive control of domestic heat pump,'' in \emph{2013 American Control Conference}.\hskip 1em plus 0.5em minus 0.4em\relax IEEE, 2013.

\bibitem{mugnini2024model}
A.~Mugnini, M.~Evens, and A.~Arteconi, ``Model predictive controls for residential buildings with heat pumps: Experimentally validated archetypes to simplify the large-scale application,'' \emph{Energy and Buildings}, vol. 320, p. 114632, 2024.

\bibitem{touretzky2014integrating}
C.~R. Touretzky and M.~Baldea, ``Integrating scheduling and control for economic mpc of buildings with energy storage,'' \emph{Journal of Process Control}, vol.~24, no.~8, pp. 1292--1300, 2014.

\bibitem{tarragona2022analysis}
J.~Tarragona, A.~L. Pisello, C.~Fern{\'a}ndez, L.~F. Cabeza, J.~Pay{\'a}, J.~Marchante-Avellaneda, and A.~de~Gracia, ``Analysis of thermal energy storage tanks and pv panels combinations in different buildings controlled through model predictive control,'' \emph{Energy}, vol. 239, p. 122201, 2022.

\bibitem{kinab2010reversible}
E.~Kinab, D.~Marchio, P.~Rivi{\`e}re, and A.~Zoughaib, ``Reversible heat pump model for seasonal performance optimization,'' \emph{Energy and Buildings}, vol.~42, no.~12, pp. 2269--2280, 2010.

\bibitem{shin2004numerical}
M.-S. Shin, H.-S. Kim, D.-S. Jang, S.-N. Lee, Y.-S. Lee, and H.-G. Yoon, ``Numerical and experimental study on the design of a stratified thermal storage system,'' \emph{Applied thermal engineering}, vol.~24, no.~1, pp. 17--27, 2004.

\bibitem{BAETEN2016217}
B.~Baeten, T.~Confrey, S.~Pecceu, F.~Rogiers, and L.~Helsen, ``A validated model for mixing and buoyancy in stratified hot water storage tanks for use in building energy simulations,'' \emph{Applied Energy}, vol. 172, pp. 217--229, 2016.

\bibitem{vrettos2016robust}
E.~Vrettos, F.~Oldewurtel, and G.~Andersson, ``Robust energy-constrained frequency reserves from aggregations of commercial buildings,'' \emph{IEEE Transactions on Power Systems}, vol.~31, no.~6, pp. 4272--4285, 2016.

\bibitem{golmohamadi2021optimization}
H.~Golmohamadi, K.~G. Larsen, P.~G. Jensen, and I.~R. Hasrat, ``Optimization of power-to-heat flexibility for residential buildings in response to day-ahead electricity price,'' \emph{Energy and Buildings}, vol. 232, p. 110665, 2021.

\bibitem{de2016quantification}
R.~De~Coninck and L.~Helsen, ``Quantification of flexibility in buildings by cost curves--methodology and application,'' \emph{Applied Energy}, vol. 162, pp. 653--665, 2016.

\bibitem{d2019mapping}
F.~D’Ettorre, M.~De~Rosa, P.~Conti, D.~Testi, and D.~Finn, ``Mapping the energy flexibility potential of single buildings equipped with optimally-controlled heat pump, gas boilers and thermal storage,'' \emph{Sustainable Cities and Society}, vol.~50, p. 101689, 2019.

\bibitem{farrokhifar2021model}
M.~Farrokhifar, H.~Bahmani, B.~Faridpak, A.~Safari, D.~Pozo, and M.~Aiello, ``Model predictive control for demand side management in buildings: A survey,'' \emph{Sustainable Cities and Society}, vol.~75, p. 103381, 2021.

\bibitem{golmohamadi2022integration}
H.~Golmohamadi, K.~G. Larsen, P.~G. Jensen, and I.~R. Hasrat, ``Integration of flexibility potentials of district heating systems into electricity markets: A review,'' \emph{Renewable and Sustainable Energy Reviews}, vol. 159, p. 112200, 2022.

\bibitem{li2023unlocking}
Y.~Li, N.~Yorke-Smith, and T.~Keviczky, ``Unlocking energy flexibility from thermal inertia of buildings: A robust optimization approach,'' in \emph{2023 62nd IEEE Conference on Decision and Control (CDC)}.\hskip 1em plus 0.5em minus 0.4em\relax IEEE, 2023, pp. 2555--2562.

\bibitem{S22023}
\BIBentryALTinterwordspacing
M.~Konsman, E.~Werkman, and in~collaboration~with TC~205 WG 18~members, ``S2 white paper,'' 2023. [Online]. Available: \url{https://s2standard.org/}
\BIBentrySTDinterwordspacing

\bibitem{konsman2020unlocking}
M.~J. Konsman, W.~E. Wijbrandi, and G.~B. Huitema, ``Unlocking residential energy flexibility on a large scale through a newly standardized interface,'' in \emph{2020 IEEE Power \& Energy Society Innovative Smart Grid Technologies Conference (ISGT)}.\hskip 1em plus 0.5em minus 0.4em\relax IEEE, 2020, pp. 1--5.

\bibitem{li2023robust}
Y.~Li, N.~Yorke-Smith, and T.~Keviczky, ``Robust optimal control with binary adjustable uncertainties,'' in \emph{2024 European Control Conference (ECC)}.\hskip 1em plus 0.5em minus 0.4em\relax IEEE, 2024, pp. 3721--3727.

\bibitem{10666588}
W.~Tang, Y.~Li, S.~Walker, and T.~Keviczky, ``Model predictive control design for unlocking the energy flexibility of heat pump and thermal energy storage systems,'' in \emph{2024 IEEE Conference on Control Technology and Applications (CCTA)}, 2024, pp. 433--439.

\bibitem{correa2018air}
F.~Correa and C.~Cuevas, ``Air-water heat pump modelling for residential heating and domestic hot water in {Chile},'' \emph{Applied Thermal Engineering}, vol. 143, pp. 594--606, 2018.

\bibitem{wang1983heat}
Y.~Wang, D.~Wilson, and D.~Neale, ``Heat-pump control,'' in \emph{IEE Proceedings D (Control Theory and Applications)}, vol. 130, no.~6.\hskip 1em plus 0.5em minus 0.4em\relax IET, 1983, pp. 328--332.

\bibitem{felten2018value}
B.~Felten and C.~Weber, ``The value (s) of flexible heat pumps--assessment of technical and economic conditions,'' \emph{Applied Energy}, vol. 228, pp. 1292--1319, 2018.

\bibitem{luickx2008influence}
P.~J. Luickx, L.~M. Helsen, and W.~D. D’haeseleer, ``{Influence of massive heat-pump introduction on the electricity-generation mix and the GHG effect: Comparison between Belgium, France, Germany and The Netherlands},'' \emph{Renewable and Sustainable Energy Reviews}, vol.~12, no.~8, pp. 2140--2158, 2008.

\bibitem{DELACRUZLOREDO2023120556}
I.~{De la Cruz-Loredo}, D.~Zinsmeister, T.~Licklederer, C.~E. Ugalde-Loo, D.~A. Morales, H.~Bastida, V.~S. Perić, and A.~Saleem, ``Experimental validation of a hybrid {1-D} multi-node model of a hot water thermal energy storage tank,'' \emph{Applied Energy}, vol. 332, p. 120556, 2023.

\bibitem{rastegarpour2018predictive}
S.~Rastegarpour, M.~Ghaemi, and L.~Ferrarini, ``A predictive control strategy for energy management in buildings with radiant floors and thermal storage,'' in \emph{2018 SICE International Symposium on Control Systems (SICE ISCS)}.\hskip 1em plus 0.5em minus 0.4em\relax IEEE, 2018, pp. 67--73.

\bibitem{dubey2021study}
A.~K. Dubey, A.~Kumar, V.~Garc{\'\i}a-D{\'\i}az, A.~K. Sharma, and K.~Kanhaiya, ``Study and analysis of sarima and lstm in forecasting time series data,'' \emph{Sustainable Energy Technologies and Assessments}, vol.~47, p. 101474, 2021.

\bibitem{box2015time}
G.~E. Box, G.~M. Jenkins, G.~C. Reinsel, and G.~M. Ljung, \emph{Time series analysis: forecasting and control}.\hskip 1em plus 0.5em minus 0.4em\relax John Wiley \& Sons, 2015.

\bibitem{junker2018characterizing}
R.~G. Junker, A.~G. Azar, R.~A. Lopes, K.~B. Lindberg, G.~Reynders, R.~Relan, and H.~Madsen, ``Characterizing the energy flexibility of buildings and districts,'' \emph{Applied Energy}, vol. 225, pp. 175--182, 2018.

\bibitem{soren17}
S.~{\O}. Jensen, A.~Marszal-Pomianowska, R.~Lollini, W.~Pasut, A.~Knotzer, P.~Engelmann, A.~Stafford, and G.~Reynders, ``{IEA EBC} annex 67 energy flexible buildings,'' \emph{Energy and Buildings}, vol. 155, pp. 25--34, 2017.

\bibitem{hart2011pyomo}
W.~E. Hart, J.-P. Watson, and D.~L. Woodruff, ``Pyomo: modeling and solving mathematical programs in python,'' \emph{Mathematical Programming Computation}, vol.~3, no.~3, pp. 219--260, 2011.

\bibitem{gurobi}
\BIBentryALTinterwordspacing
{Gurobi Optimization, LLC}, \emph{{Gurobi Optimizer Reference Manual}}, 2024. [Online]. Available: \url{https://www.gurobi.com}
\BIBentrySTDinterwordspacing

\bibitem{pandas_api}
\BIBentryALTinterwordspacing
{Pandas Development Team}, \emph{{pandas: API Reference}}, 2024, version 2.2.3. [Online]. Available: \url{https://pandas.pydata.org/pandas-docs/stable/reference/api/pandas.DataFrame.interpolate.html}
\BIBentrySTDinterwordspacing

\bibitem{schwickart2016flexible}
T.~Schwickart, H.~Voos, M.~Darouach, and S.~Bezzaoucha, ``A flexible move blocking strategy to speed up model-predictive control while retaining a high tracking performance,'' in \emph{2016 European Control Conference (ECC)}.\hskip 1em plus 0.5em minus 0.4em\relax IEEE, 2016, pp. 764--769.

\end{thebibliography}

\end{document}